\documentclass[twocolumn]{article}

\usepackage{arxiv}

\usepackage{biblatex}
\usepackage{enumitem}
\usepackage{svg}
\usepackage[utf8]{inputenc}
\usepackage[T1]{fontenc}
\usepackage{url}
\usepackage[colorlinks=false]{hyperref}
\usepackage{booktabs}
\usepackage{amsfonts}
\usepackage{nicefrac}
\usepackage{microtype}
\usepackage{subcaption}
\usepackage{graphicx}
\usepackage{doi}
\usepackage{listings}
\usepackage{xcolor}
\usepackage{algorithm}
\usepackage{algorithmicx}
\usepackage{algpseudocode}
\hypersetup{
    citecolor = [rgb]{0,0,0.5},
}

\lstdefinestyle{mystyle}{
    backgroundcolor=\color{backcolour},   
    commentstyle=\color{codegreen},
    keywordstyle=\color{magenta},
    numberstyle=\tiny\color{codegray},
    stringstyle=\color{codepurple},
    basicstyle=\ttfamily\footnotesize,
    breaklines=true,
    captionpos=b,
    keepspaces=true,
    numbers=left,
    numbersep=5pt,
    showspaces=false,
    showstringspaces=false,
    showtabs=false,
    tabsize=2
}

\title{Challenging GPU Dominance: When CPUs Outperform for On-Device LLM Inference}

\author{
    {\hspace{1mm}Haolin Zhang} \\
    Department of Computer Science \& Engineering\\
    Texas A\&M University\\
    College Station, TX 77843 \\
    \texttt{chris\_zhang@tamu.edu} \\
    \And
    {\hspace{1mm}Jeff Huang} \\
    Department of Computer Science \& Engineering\\
    Texas A\&M University\\
    College Station, TX 77843 \\
    \texttt{jeffhuang@tamu.edu} \\
}

\date{}
\renewcommand{\headeright}{}
\renewcommand{\undertitle}{}

\hypersetup{
pdftitle={Challenging GPU Dominance: When CPUs Outperform for On-Device LLM Inference},
pdfsubject={Machine Learning, System, Tech Report},
pdfauthor={Haolin Zhang, Jeff Huang},
pdfkeywords={On-Device AI, LLM Inference, Optimization, ARM Architecture, Mobile Machine Learning},
}

\begin{document}

\twocolumn[ 
    \maketitle

    \begin{abstract}
    The common assumption in on-device AI is that GPUs, with their superior parallel processing, always provide the best performance for large language model (LLM) inference. In this work, we challenge this notion by empirically demonstrating that, under certain conditions, CPUs can outperform GPUs for LLM inference on mobile devices. Using a 1-billion-parameter LLM deployed via llama.cpp on the iPhone 15 Pro, we show that a CPU-only configuration (two threads, F16 precision) achieves 17 tokens per second, surpassing the 12.8 tokens per second obtained with GPU acceleration. We analyze the architectural factors driving this counterintuitive result, revealing that GPU memory transfer overhead and CPU thread optimization play a critical role. Furthermore, we explore the impact of thread oversubscription, quantization strategies, and hardware constraints, providing new insights into efficient on-device AI execution. Our findings challenge conventional GPU-first thinking, highlighting the untapped potential of optimized CPU inference and paving the way for smarter deployment strategies in mobile AI. However, fully explaining the observed CPU advantage remains difficult due to limited access to low-level profiling tools on iOS.
    \end{abstract}
    
    \vspace{1cm} 
]


\section{Introduction}

Recent advances in \emph{small language models (SLMs)} have challenged the conventional focus on scaling up AI models. While GPUs have been the dominant choice for inference, emerging evidence suggests that they may not always offer the best performance, particularly for SLMs on mobile devices \cite{Czerski_2025}. This shift is reshaping AI deployment, as compact models demonstrate high efficiency across diverse tasks.

Models such as Phi \cite{abdin2024phi, javaheripi2023phi}, LLaMA 3.2 Herd \cite{grattafiori2024llama3herdmodels}, TinyLLaMA \cite{zhang2024tinyllamaopensourcesmalllanguage}, Apple Foundation Models \cite{gunter2024appleintelligencefoundationlanguage}, and Qwen-2 \cite{yang2024qwen2technicalreport} exemplify this trend. They show that smaller architectures can achieve strong performance in natural language processing, expanding the possibilities for real-world applications beyond large-scale cloud infrastructure.

Deploying SLMs on mobile devices presents unique challenges due to constraints in memory, processing power, and energy consumption. Researchers have proposed optimizations to address these limitations. Liu et al. \cite{liu2024mobilellm} introduced MobileLLM, an architecture for sub-billion parameter models that optimizes embedding sharing, grouped-query attention, and deep-thin structures to minimize resource overhead. Pham et al. \cite{pham2024slimlmefficientsmalllanguage} developed SlimLM, fine-tuned for document assistance on smartphones, balancing model size, context length, and inference speed. Li et al. \cite{li2024transformerlitehighefficiencydeploymentlarge} presented Transformer-Lite, an inference engine that optimizes mobile GPU efficiency with dynamic shape support, FP4 quantization, and sub-tensor KV cache handling, achieving notable speedups.

These advancements have fueled interest in on-device SLM deployment, reducing reliance on cloud services, improving privacy, and lowering latency and costs. A key tool in this domain is llama.cpp \cite{llama.cpp}, an open-source framework optimized for consumer hardware, including mobile devices. It supports various performance optimizations, such as quantization \cite{kquants} and multi-threaded execution, making it a valuable platform for pushing the limits of on-device AI.

This paper examines the performance of SLMs on mobile devices. We evaluate six models, ranging from 0.5 billion to 8 billion parameters, under different precision settings to assess the impact of quantization and hardware configurations on inference speed.

Given the iPhone’s prominence in the premium market and Apple’s growing investment in on-device AI \cite{appleintelligence}, we selected the iPhone~15~Pro as our test platform. It features an Apple-designed A17 Pro SoC with an ARM v8.6-A ISA and a Metal-supported GPU. We use llama.cpp as the inference engine for our experiments.

Our study highlights three key findings:

\begin{itemize}[leftmargin=0pt]
    \item \textbf{CPU Competitiveness with GPU:} Under specific conditions, \emph{CPU-only} inference can match or even surpass GPU performance. For smaller models such as Qwen2-0.5B and LLaMA-3.2-1B, multi-threaded CPU execution with \texttt{Q4\footnote{Effective 4.5 bits/weight}} and \texttt{F16} precision achieves speedups of 1.31$\times$ and 1.33$\times$, respectively, over GPU execution. However, for models larger than 1.5B parameters, GPUs generally maintain higher throughput, as seen with Mistral-7B and LLaMA-3.2-8B, where CPU performance declines due to increased memory and computation demands.
    
    \item \textbf{Matrix Multiplication as the Bottleneck:} Profiling reveals that matrix multiplication (\texttt{GGML\_OP\_MUL\_MAT}) dominates computation, accounting for 87.6\% of execution time in the prefill phase and 76.2\% in the decode phase\footnote{For llama3.2-1B@f16}. This underscores the need to optimize General Matrix Multiplication (GEMM) operations to improve SLM inference efficiency.
    
    \item \textbf{Impact of Thread Allocation:} Optimal performance occurs when the number of CPU threads matches the architecture’s performance cores. Adding more threads beyond this threshold yields diminishing returns or performance degradation, emphasizing the importance of architecture-aware thread scheduling.
\end{itemize}

Through these insights, we challenge the dominant “GPU-first” paradigm, demonstrating that CPUs can be competitive for on-device AI. Our findings contribute to the development of resource-efficient, high-performance AI systems. 

The rest of the paper is organized as follows. Section~\ref{Background} describes the llama.cpp compute graph and its role in model inference. Section~\ref{Result} presents a performance analysis of models from 0.5B to 8B parameters across different backends and precision settings. Section ~\ref{Discussion} discussed profiling results from llama.cpp runtime. Section~\ref{Topological-Based Graph Execution} proposes a hardware-aware execution strategy to maximize concurrency. Finally, Sections~\ref{Limitations}--\ref{Conclusion} discuss profiling challenges on mobile hardware and future directions for optimizing both CPU and GPU performance.

\section{Related Work} \label{Related Work}

\subsection{Advancements in Small Language Models}
Research on \emph{small language models} (SLMs) has aimed to achieve strong performance with lower computational costs, making them practical for mobile and edge applications. Several models have been introduced, including Meta’s OPT \cite{zhang2022optopenpretrainedtransformer} and LLaMA series \cite{touvron2023llamaopenefficientfoundation, grattafiori2024llama3herdmodels}, Microsoft’s Phi series \cite{javaheripi2023phi, abdin2024phi}, the Qwen series \cite{yang2024qwen2technicalreport}, StabilityAI’s StableLM \cite{bellagente2024stablelm216b}, and Google's Gemma \cite{gemmateam2024gemmaopenmodelsbased} and Gemini Nano \cite{geminiteam2024geminifamilyhighlycapable}. Other models, such as MobileVLM \cite{chu2023mobilevlmfaststrong} and Apple Foundation Models \cite{gunter2024appleintelligencefoundationlanguage}, have been designed with efficiency as a priority for resource-limited environments.

One key improvement in modern SLMs is the refinement of self-attention mechanisms, which traditionally scale quadratically with sequence length. Methods such as \emph{flash attention} \cite{dao2022flashattention} optimize memory use by reducing redundant operations, improving inference speed. \emph{Grouped-query attention (GQA)} \cite{ainslie2023gqatraininggeneralizedmultiquery}, used in the Qwen and LLaMA 3.2 series, reduces memory overhead by processing queries in groups, offering efficiency without sacrificing accuracy.

Compression methods such as quantization further improve inference by reducing memory and bandwidth needs. Low-bit approaches, including GPTQ \cite{frantar2023gptqaccurateposttrainingquantization} and llama.cpp's k-quant \cite{kquants}, minimize storage and computational requirements while maintaining precision. Recent work, such as SmoothQuant \cite{xiao2024smoothquantaccurateefficientposttraining}, extends these benefits by introducing weight-activation co-quantization to lower computational load while maintaining stable outputs.

These developments improve the balance between efficiency and accuracy, making smaller models a practical alternative for on-device use. However, the efficiency of CPUs compared to GPUs for running these models under different quantization settings and model sizes is still not well understood. This study addresses this question by systematically evaluating CPU-GPU performance trade-offs across multiple configurations.

\subsection{On-Device Deployment of SLMs}
Efficient on-device inference has led to the development of specialized engines, each with different optimization strategies. MNN \cite{jiang2020mnnuniversalefficientinference} offers a highly optimized mobile inference engine with kernel optimizations and hybrid backend scheduling. However, its execution model focuses on individual operators rather than full compute graph scheduling, which affects performance on mobile devices with heterogeneous computing architectures. MNN also does not support quantization below 8-bit, limiting its effectiveness for larger models in memory-constrained environments.

PowerInfer \cite{song2024powerinferfastlargelanguage} improves performance by distributing frequently activated neurons on the GPU while offloading others to the CPU. This approach enhances execution speed on consumer GPUs and supports models as large as OPT-175B. However, the method is designed for desktop environments, which limits its direct applicability to mobile and embedded devices.

ExecuTorch \cite{executorch}, an extension of PyTorch, broadens compatibility with edge devices while prioritizing ease of use and hardware efficiency across CPUs, NPUs, and DSPs. It relies on the PyTorch 2 compiler for model export, addressing some limitations of previous approaches, such as TorchScript in PyTorch Mobile. However, balancing a small memory footprint with execution speed remains an ongoing challenge.

MediaPipe \cite{lugaresi2019mediapipeframeworkbuildingperception} is widely used for vision-based applications and provides a modular framework for evaluating system performance across platforms. While it allows for rapid prototyping, adapting it for transformer-based models may require modifications.

Recent studies explore extreme quantization methods, such as BitNet b1.58 \cite{ma2024era1bitllmslarge}, which reduces model weights to ternary values (-1, 0, 1) while maintaining performance close to full-precision models. This method significantly reduces latency, memory use, and energy consumption. BitNet b1.58 also introduces a new training approach that supports efficient model scaling and informs future hardware design for low-precision inference.

These inference engines and compression methods highlight different strategies for deploying LLMs on mobile and embedded devices, from hardware-specific optimizations to more flexible, cross-platform solutions.

\subsection{Measurement and Survey of On-Device LLM Deployment}
The study of language model performance on resource-constrained devices has gained attention, leading to systematic evaluations. Lu et al. \cite{lu2024smalllanguagemodelssurvey} provide a survey of SLMs ranging from 100M to 5B parameters, covering architecture, training methods, and key performance metrics, such as inference speed and memory use. Their analysis focuses on transformer-based, decoder-only models common in smart devices. Li et al. \cite{li2024llminferenceservingsurvey} examine inference serving optimizations that improve efficiency without modifying core decoding methods.

Xu et al. \cite{xu2024ondevicelanguagemodelscomprehensive} review the challenges of running computationally intensive models on edge devices. Their study covers efficient architectures, compression techniques (quantization, pruning, knowledge distillation), and hardware acceleration methods, all critical for balancing accuracy with efficiency. Li et al. \cite{li2024llmonmobiledevices} evaluate 22 LLMs across four mobile devices, comparing accuracy, latency, and memory usage. Their results suggest that mobile inference engines often show little advantage from GPU acceleration, with some cases demonstrating slower performance. They also find that smaller models, such as Bloom 0.5B, achieve a 10.5$\times$ speedup over Bloom 7B with only a 1.37\% accuracy reduction, supporting the idea that SLMs can provide efficient inference with minimal loss of precision.

This study builds on previous research by providing an in-depth analysis of the llama.cpp inference framework, including its technical improvements. A new execution approach based on compute graph order is introduced, designed to optimize hardware usage and mitigate performance inefficiencies found in earlier work. Unlike prior studies that focus primarily on profiling mobile LLM performance, this research examines the deployment of 1-billion-parameter models on physical devices, offering insights for practical applications.

\section{Background} \label{Background}

llama.cpp\footnote{Code based on commit 8648c52} represents machine learning models as compute graphs, where each node corresponds to an operation in the model architecture. Model files are stored in the \texttt{gguf} format, containing both the model type (LLaMA, Baichuan, Falcon, Grok, etc.) and weight parameters. When a model is loaded, llama.cpp calls the \texttt{llama\_build\_graph} function, which dynamically constructs the compute graph based on the model architecture.

For models in the LLaMA family, this graph is assembled using the \texttt{build\_llama()} function. Algorithm~\ref{alg:build_llama} outlines the process of creating transformer layers, including attention mechanisms, feed-forward networks, and residual connections.

\begin{algorithm}[h]
\caption{Pseudocode for \texttt{build\_llama}}
\label{alg:build_llama}
\begin{algorithmic}[1]
\Function{build\_llama}{}
    \State \Call{init}{gf, inpL}
    \For{each layer}
        \State norm\_inp $\gets$ \Call{norm}{inpL}
        \State $Q, K, V \gets$ \Call{attn\_weights}{norm\_inp}
        \State $Q, K \gets$ \Call{rotary}{Q, K}
        \State attn\_out $\gets$ \Call{attention}{Q, K, V}
        \State ffn\_inp $\gets$ \Call{add}{attn\_out, inpL}
        \State ffn\_norm $\gets$ \Call{ffn\_norm}{ffn\_inp}
        \State ffn\_out $\gets$ \Call{ffn}{ffn\_norm}
        \State inpL $\gets$ \Call{add}{ffn\_out, inpL}
    \EndFor
    \State final\_norm $\gets$ \Call{norm}{inpL}
    \State final\_out $\gets$ \Call{apply\_weight}{final\_norm}
    \State \Call{build\_forward\_pass}{gf, final\_out}
    \State \Return gf
\EndFunction
\end{algorithmic}
\end{algorithm}

\begin{figure*}[!t]
    \centering
    \includegraphics[width=\textwidth]{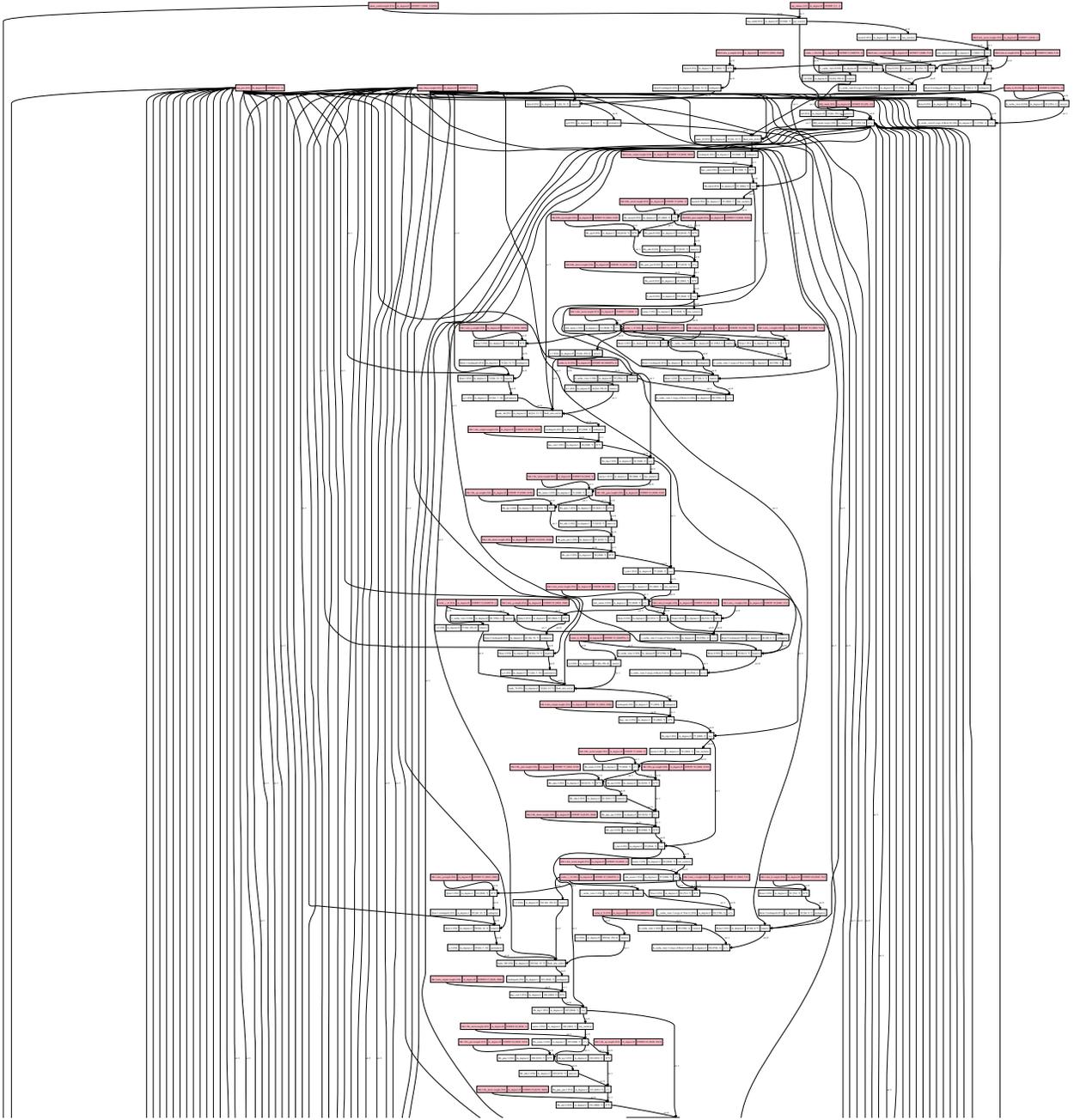}
    \caption{Segment of the compute graph for a LLaMA 3.2 model.}
    \label{fig:compute_graph_example}
\end{figure*}

This function builds the compute graph to match the hierarchical structure of the LLaMA transformer model. Figure~\ref{fig:compute_graph_example} shows a segment of the compute graph for \textbf{LLaMA 3.2-1B}, illustrating the structure of decoder blocks. The full graph can be accessed on \href{https://github.com/Chrisz236/llama.cpp_tech_report/tree/main/graphs}{GitHub}.

After constructing the graph, execution is managed by \textit{ggml}, a tensor library optimized for different hardware backends. In llama.cpp, execution follows a sequential schedule, where each node—representing an operation such as matrix multiplication or normalization—is processed in order based on its dependencies.

\begin{figure}[htp]
    \centering
    \includegraphics[width=\columnwidth]{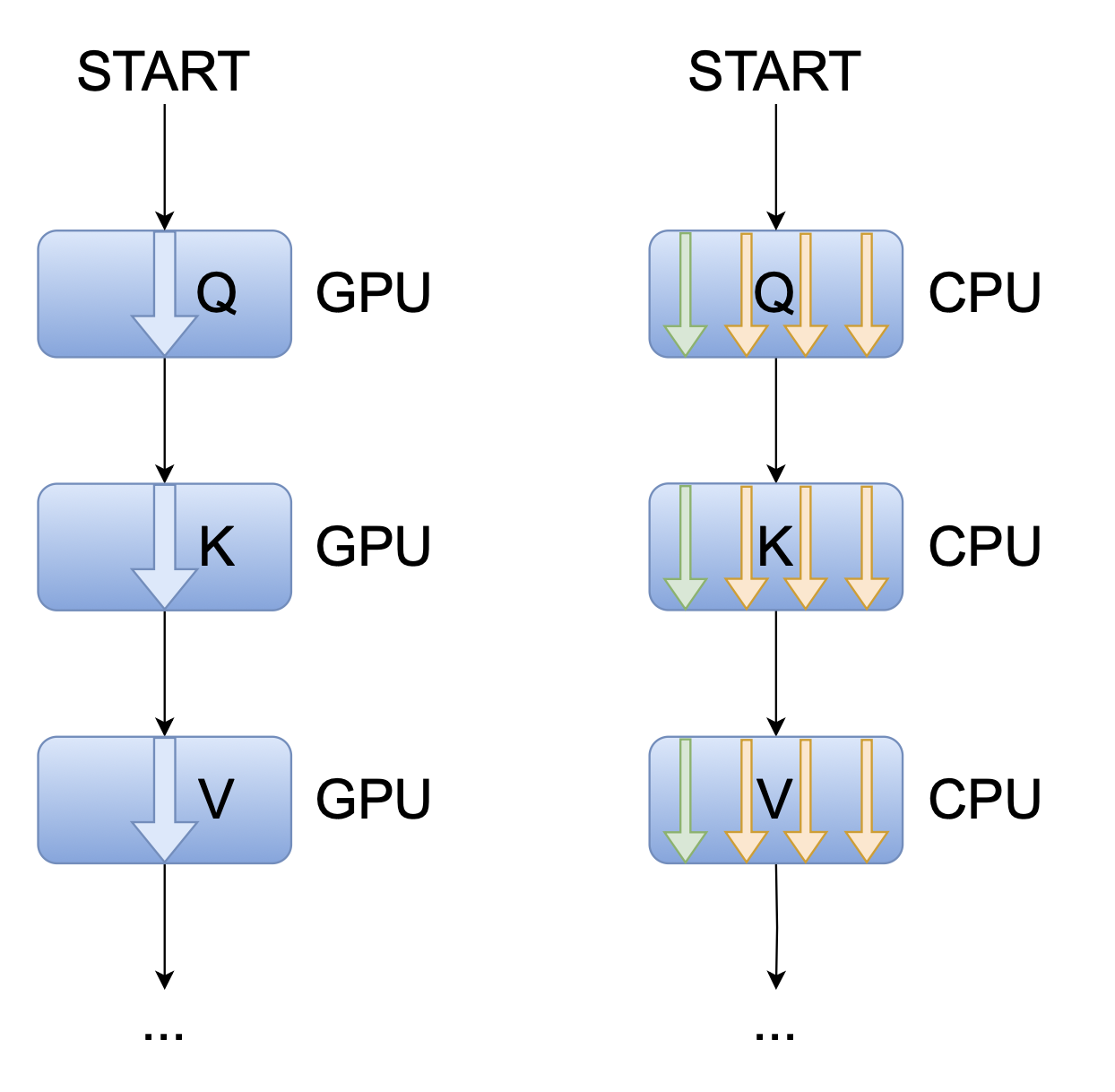}
    \caption{\textbf{Left (GPU-enabled)}: Nodes are offloaded to the GPU backend. 
    \textbf{Right (CPU-only execution)}: Nodes are processed in parallel across four CPU threads.}
    \label{fig:llama_cpp_default}
\end{figure}

Figure~\ref{fig:llama_cpp_default} shows the execution schedule for both GPU-enabled and CPU-only configurations. When GPU acceleration is available, nodes are offloaded to the GPU backend. In a CPU-only setup, computations are parallelized across multiple threads.

\begin{figure*}[!t]
    \centering
    \includegraphics[width=\textwidth]{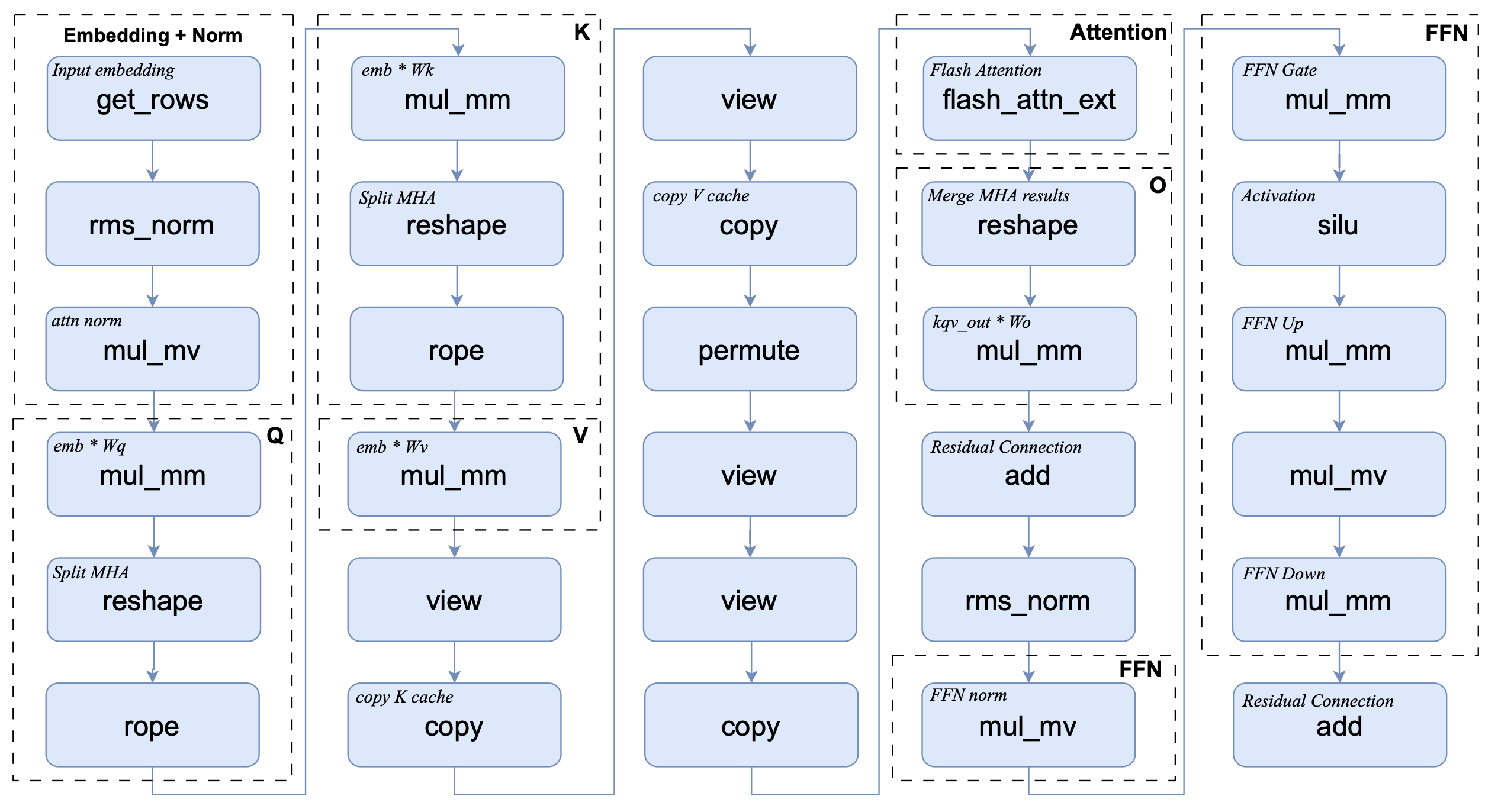}
    \caption{Execution sequence of nodes within transformer blocks. Each operation is processed sequentially.}
    \label{fig:serial_execution}
\end{figure*}

Even with parallel execution at the hardware level, operations within each transformer block are scheduled \textbf{serially}, meaning attention, residual connections, and feed-forward computations are executed one after another within the same layer. Figure~\ref{fig:serial_execution} illustrates this ordered execution, showing how dependencies are handled in the compute graph.

Each node represents a fundamental mathematical operation, such as \textbf{matrix multiplication}, \textbf{element-wise addition}, or \textbf{rotary position encoding (ROPE)}. The execution backend is selected at compile time, based on hardware capabilities and compiler settings. Supported backends include \texttt{NEON}, \texttt{AVX/AVX2}, \texttt{RISC-V Intrinsic}, \texttt{Power9 Vector}, \texttt{Loongarch ASX} for CPU acceleration, and \texttt{CUDA}, \texttt{MUSA} \texttt{HIP}, \texttt{Vulkan}, \texttt{SYCL} and \texttt{Metal} for GPU execution. This flexibility allows llama.cpp to achieve efficient performance across different computing platforms.

\section{Methodology} \label{Methodology}

\subsection{Hardware and Software Setup}
To evaluate the feasibility of running large language models on mobile devices, we conducted experiments on an iPhone 15 Pro, equipped with a 2-performance-core, 4-efficiency-core (2P+4E) CPU configuration and 8GB of RAM. The GPU was tested under its default enabled setting. The device operated on iOS 18.2.1, and all experiments were executed on a codebase built from commit \texttt{8648c52}, ensuring strict reproducibility. To mitigate performance variability caused by thermal throttling, we stabilized the device temperature by placing it in an ice-cooled environment.

\subsection{Model Selection and Precision}
Considering the constrained computational and memory resources of mobile devices, our study evaluates a range of small-scale large language models (LLMs) optimized for efficiency without compromising performance. We conduct experiments on models with parameter sizes of 0.5B, 1B, 1.5B, 3B, 7B, and 8B, encompassing popular architectures such as Qwen(\cite{yang2024qwen2technicalreport}), LLaMA (\cite{grattafiori2024llama3herdmodels}), and Mistral(\cite{jiang2023mistral7b}) families. To investigate the trade-offs between precision and efficiency, we evaluate these models under various precision configurations, including full-precision (F16), 8-bit quantization (Q8), and 4-bit quantization (Q4).

\subsection{Test Configurations}
To systematically analyze inference performance, we tested two execution modes:
\begin{itemize}
    \item \textbf{GPU-Enabled (Default)}: The model utilized the iPhone’s integrated GPU for acceleration, leveraging available hardware optimizations.
    \item \textbf{CPU-Only Execution}: The model executed exclusively on the CPU, with thread parallelism ranging from 1 to 6 threads to assess performance scaling.
\end{itemize}

\subsection{Experimental Procedure}
Each configuration was benchmarked using llama.cpp, measuring inference speed in tokens per second (tk/s). To ensure statistical reliability, each setting was evaluated over five independent runs, and the average throughput was recorded. To eliminate prompt variability, all tests used a fixed input: "\texttt{The meaning of life is}" (7 tokens in total).

\subsection{Evaluation Metrics}
The primary performance metric was inference throughput, defined as the number of tokens generated per second (tk/s). This metric provides a direct measure of computational efficiency across different hardware configurations and precision formats.

\section{Results} \label{Result}

\subsection{Performance Comparison Across Models and Backends}
We benchmark six models---Qwen2-0.5B, Qwen2-1.5B, Llama-3.2-1B, Llama-3.2-3B, Mistral-7B-v0.1, and Llama-3.2-8B---under various numerical precision setups (F16, Q4, and Q8 when applicable) and across diverse backends (GPU or 1--6 CPU threads). For each configuration, we measure inference speed in tokens per second (tk/s). 
Figure~\ref{fig:model_performance} presents the consolidated results, with sub-figures illustrating performance trends for each model.

\paragraph{Memory and Timeout Constraints.}
For larger models with 7B and 8B parameters, the F16 and Q8 configurations exceeded the device's memory capacity, resulting in \texttt{mmap} failures. Additionally, attempting to run 7B or 8B models using a single or dual CPU thread frequently led to timeouts, as the context windows (128 tokens) could not be filled within 1 minute. This resulted in missing data points in certain plots.

\subsection{F16 Precision Performance Analysis}
Figures~\ref{fig:model_performance}(a)--(d) illustrate the F16 performance for Qwen2-0.5B, Qwen2-1.5B, Llama-3.2-1B, and Llama-3.2-3B. Notably, for smaller-sized models (e.g., Qwen2-0.5B in Figure~\ref{fig:model_performance}(a)), CPU execution can \emph{exceed} GPU performance when multiple threads are used. For instance, on Qwen2-0.5B at F16, the GPU baseline lags behind a 2-thread CPU configuration, underscoring the efficiency of multi-threading for sub-1B parameter networks. A similar trend emerges for Llama-3.2-1B in Figure~\ref{fig:model_performance}(b), where CPU performance scales favorably up to four or five threads before diminishing returns set in.  

For mid-sized models (e.g., Qwen2-1.5B, Llama-3.2-3B), GPU acceleration generally offers higher throughput. However, the CPU remains competitive given enough threads (four or more), indicating that careful threading can offset GPU kernel launch overheads and memory-transfer bottlenecks. For 8B-F16, no results are shown in Figure~\ref{fig:model_performance}(f) due to memory allocation failures on the test device.

\subsection{Q4 (and Q8) Quantization Performance Analysis}
Quantization to 4 bits significantly boosts inference speed across all models where it is supported. In Figures~\ref{fig:model_performance}(a)--(d), we observe that moving from F16 to Q4 yields anywhere from a 1.5$\times$ to 2.5$\times$ speedup, most prominently on larger CPU thread counts. As shown in Figure~\ref{fig:model_performance}(b) for Llama-3.2-1B at Q4, the GPU baseline improves substantially over F16, while multi-threaded CPU execution narrows the performance gap with a suitable number of threads.  
For 7B and 8B models, Q8 configurations could not be fully benchmarked (memory allocation failures), and 1--2 CPU threads timed out for Q4. Nonetheless, the available data in Figures~\ref{fig:model_performance}(e) and~\ref{fig:model_performance}(f) for Mistral-7B-v0.1 (Q4) and Llama-3.2-8B (Q4) suggest that performance scales best beyond two CPU threads, though the GPU remains superior when it can be utilized.

\subsection{Observations and Discussion}
These experiments yield several key insights:
\begin{itemize}
    \item \textbf{CPU versus GPU trade-offs:} For small-scale models (e.g., below 1B parameters) at F16 precision, multi-threaded CPU execution \emph{can outperform} a GPU due to reduced kernel overheads and dynamic scheduling.
    \item \textbf{Thread Scalability:} Adding CPU threads boosts throughput until an optimal threshold (commonly four or five in our tests), after which resource contention and memory bandwidth limits degrade performance.
    \item \textbf{Quantization Benefits:} Q4 compression offers a noticeable increase in tokens per second across the tested backends, making it attractive for performance-sensitive applications that can tolerate small accuracy trade-offs.
    \item \textbf{Limitations on Larger Models:} For 7B and 8B models, memory constraints (F16/Q8) and execution timeouts (1--2 threads) prevent a straightforward CPU-GPU comparison. Higher-capacity hardware or further quantization would be required for real-time inference on these larger networks.
\end{itemize}

\begin{figure*}[t]  
    \centering

    \begin{subfigure}[t]{0.32\textwidth}
        \centering
        \includegraphics[width=\linewidth]{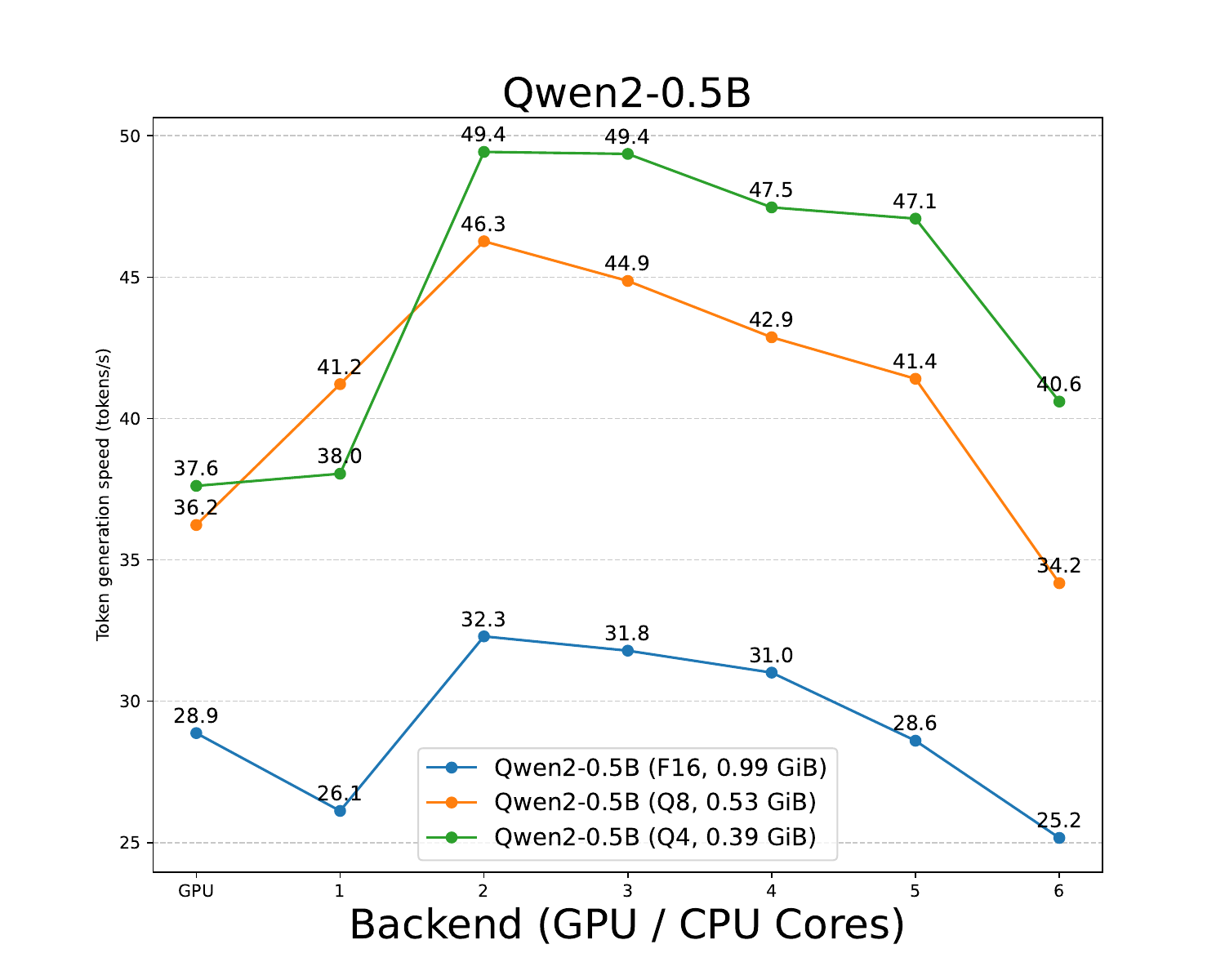}
        \caption{Qwen2-0.5B}
    \end{subfigure}
    \begin{subfigure}[t]{0.32\textwidth}
        \centering
        \includegraphics[width=\linewidth]{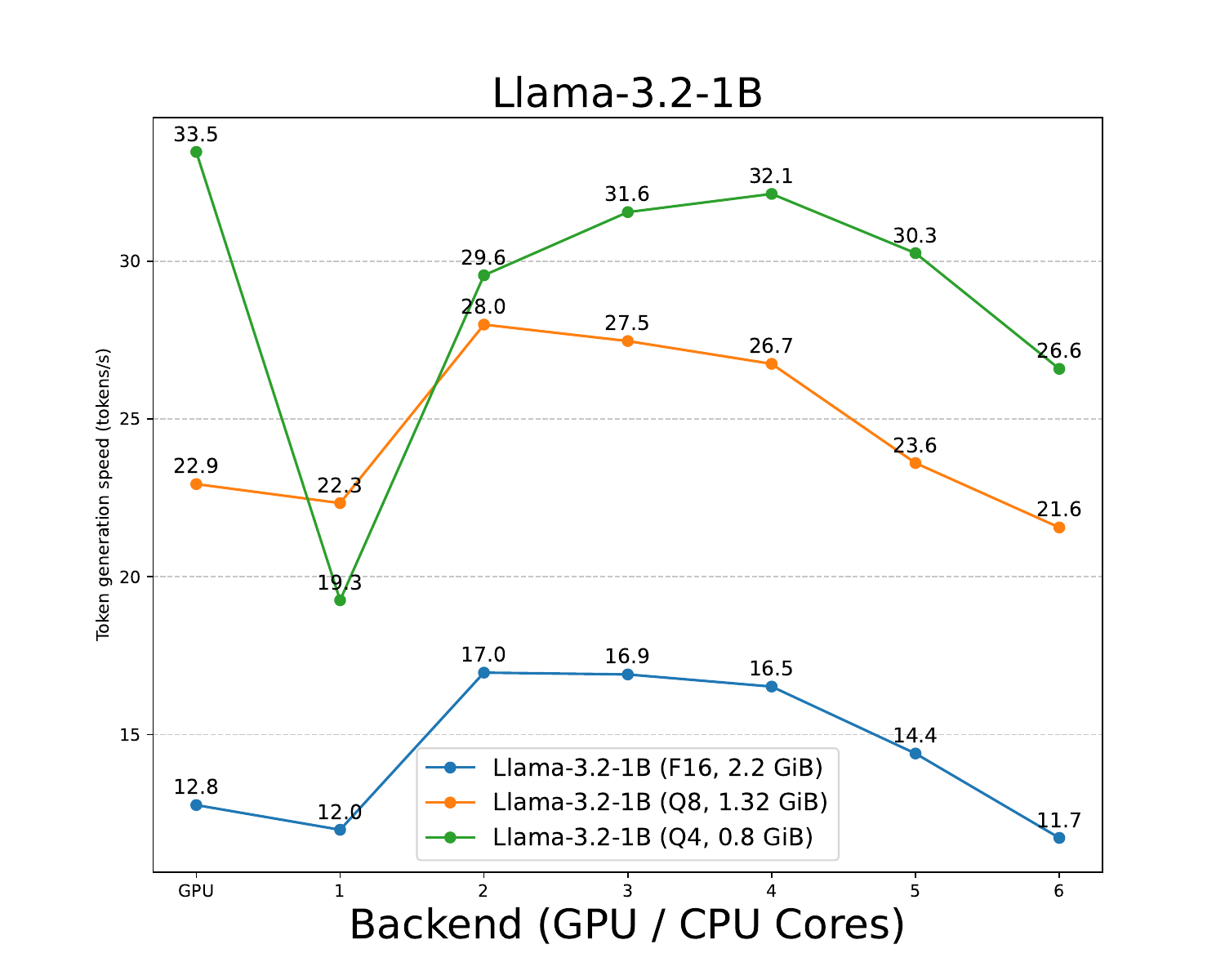}
        \caption{Llama-3.2-1B}
    \end{subfigure}
    \begin{subfigure}[t]{0.32\textwidth}
        \centering
        \includegraphics[width=\linewidth]{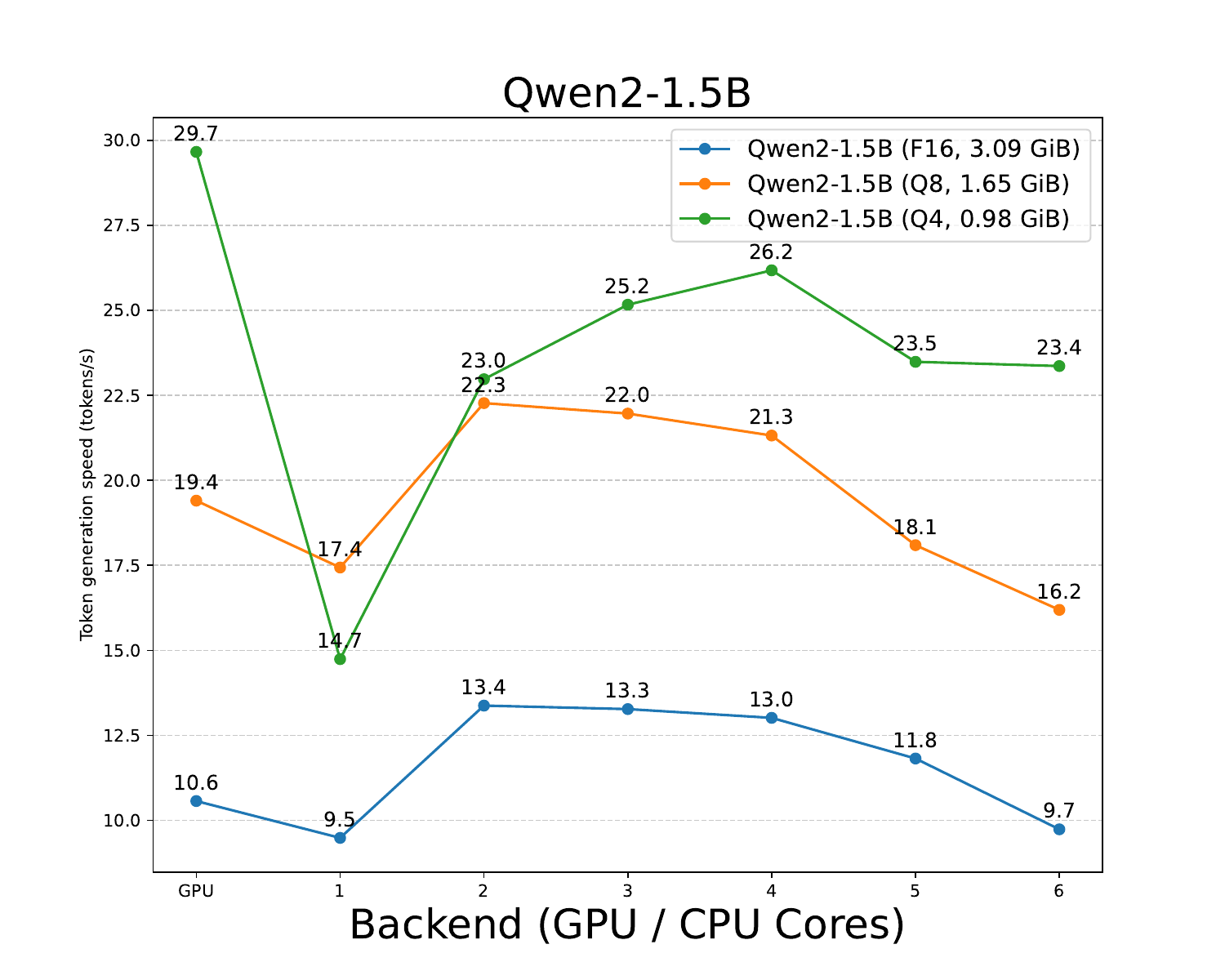}
        \caption{Qwen2-1.5B}
    \end{subfigure}

    \begin{subfigure}[t]{0.32\textwidth}
        \centering
        \includegraphics[width=\linewidth]{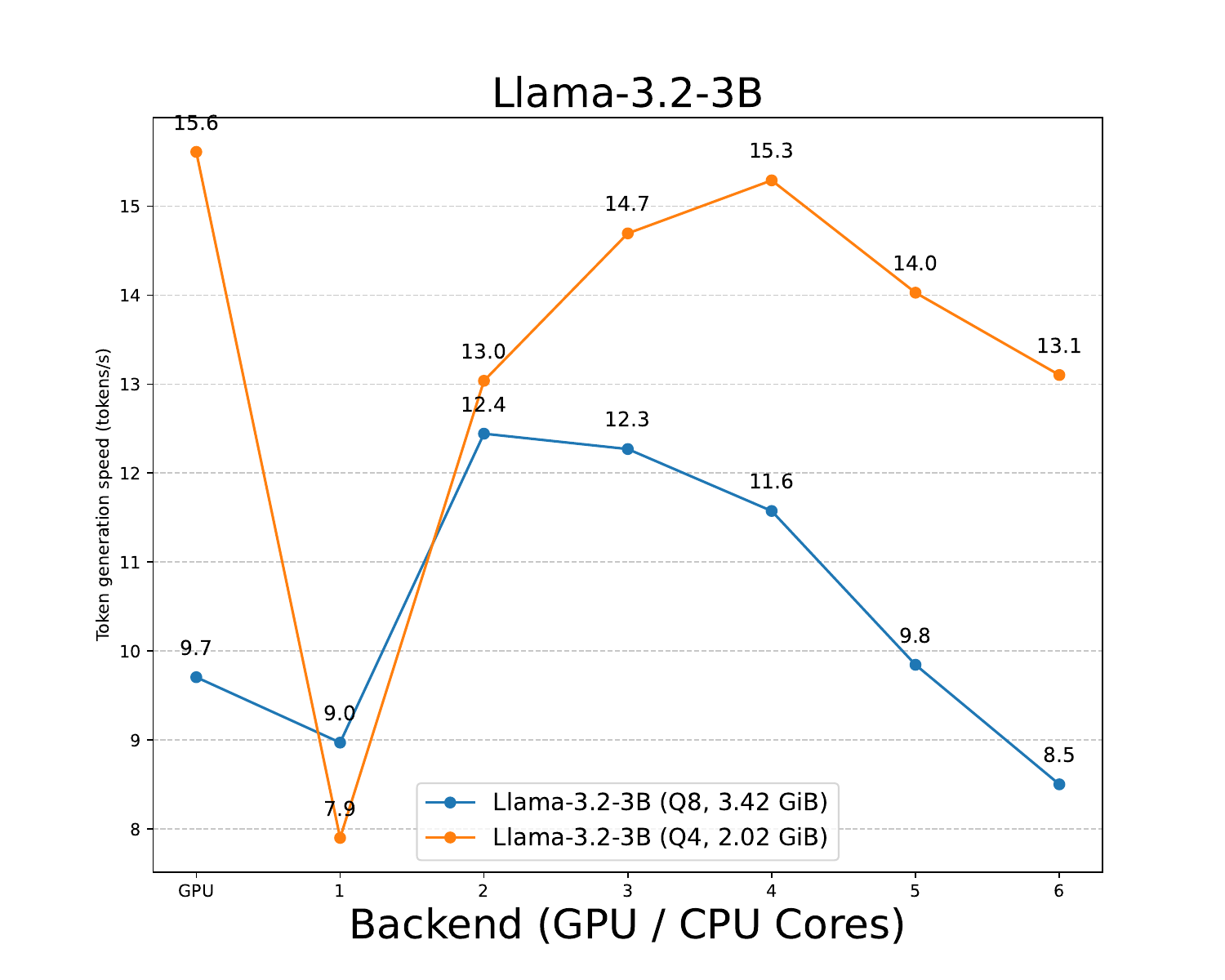}
        \caption{Llama-3.2-3B}
    \end{subfigure}
    \begin{subfigure}[t]{0.32\textwidth}
        \centering
        \includegraphics[width=\linewidth]{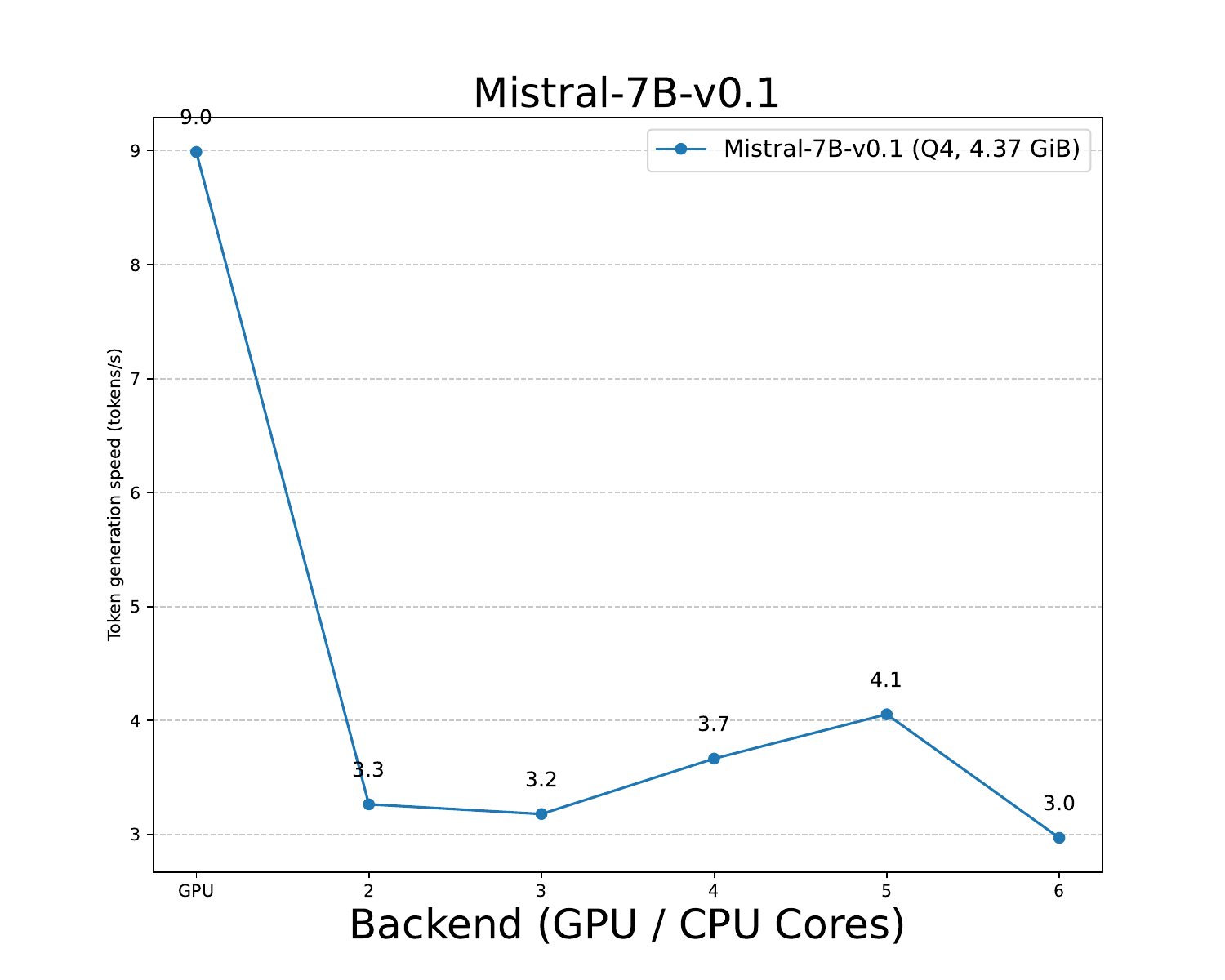}
        \caption{Mistral-7B-v0.1}
    \end{subfigure}
    \begin{subfigure}[t]{0.32\textwidth}
        \centering
        \includegraphics[width=\linewidth]{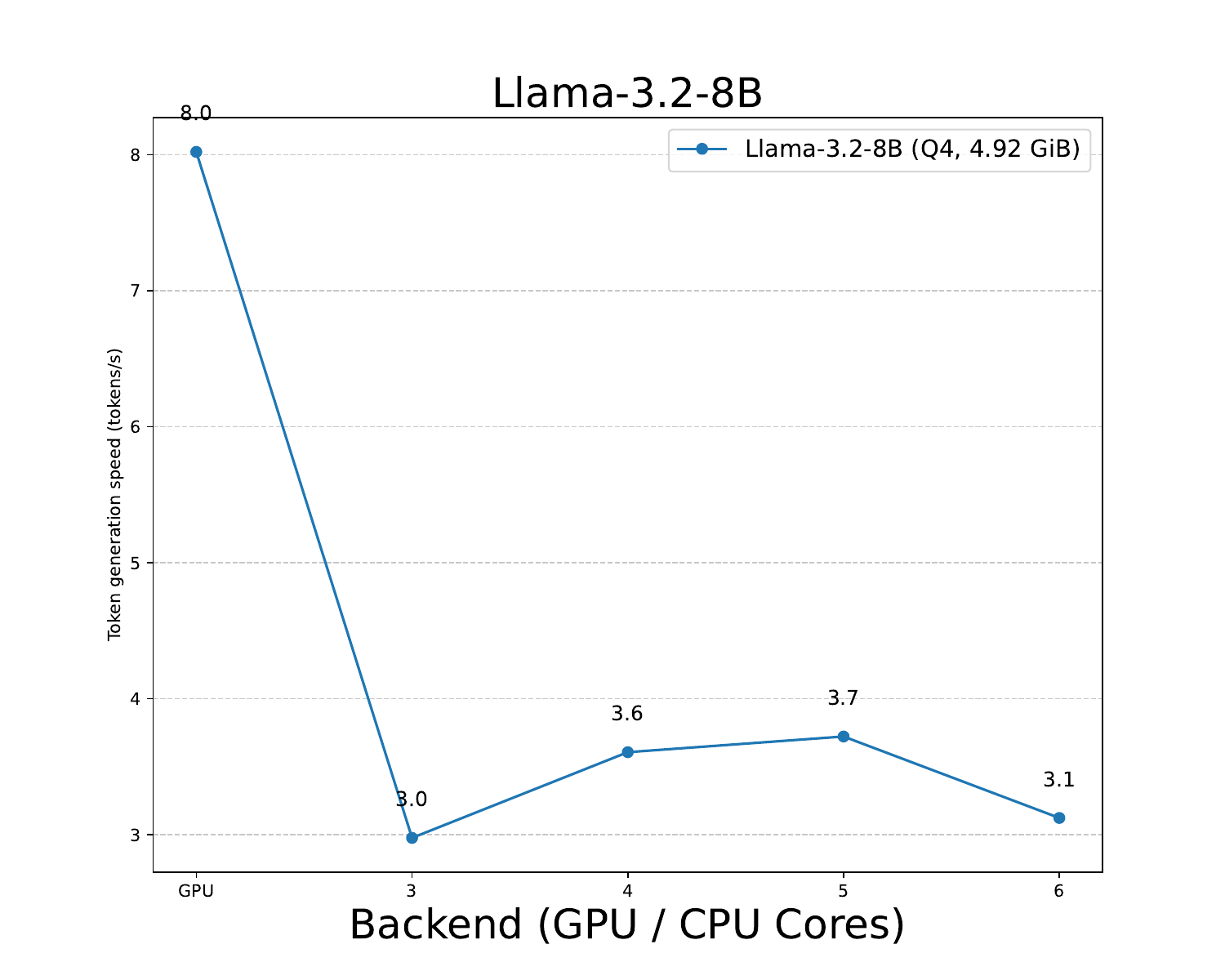}
        \caption{Llama-3.2-8B}
    \end{subfigure}

    \caption{Performance Comparison of Different Models Across Backends}
    \label{fig:model_performance}
\end{figure*}

Overall, Figure~\ref{fig:model_performance} demonstrates that careful choice of numerical precision (F16 vs.\ Q4) and backend configuration (GPU vs.\ CPU threading) is crucial for balancing model size, hardware constraints, and desired inference speed.

\textbf{Remark}: Interestingly, our observations indicated that smaller LLM inference on GPUs can exhibit lower performance compared to CPUs. This counterintuitive behavior, despite repeated verification, remains without a detailed low-level explanation. The primary reason for this gap is the absence of suitable profiling and diagnostic tools to precisely analyze system overheads, memory transfer bottlenecks, and parallelization inefficiencies on iOS devices.

\begin{figure}[htp]
    \centering
    \begin{subfigure}[b]{0.48\textwidth}
        \centering
        \includegraphics[width=\columnwidth]{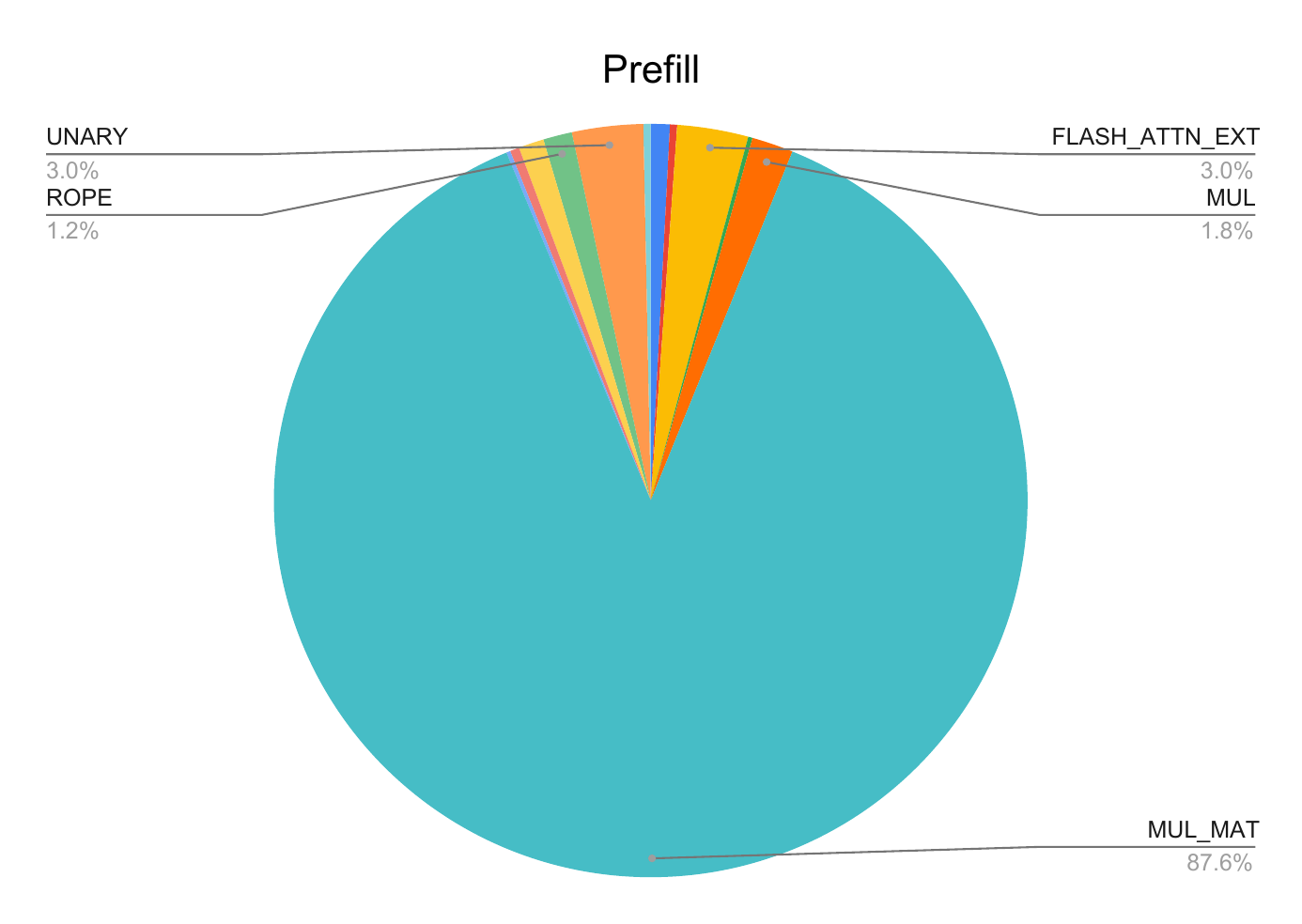}
        \caption{OPs performance in prefill stage}
        \label{fig:prefill_ops}
    \end{subfigure}
    \hfill
    \begin{subfigure}[b]{0.48\textwidth}
        \centering
        \includegraphics[width=\textwidth]{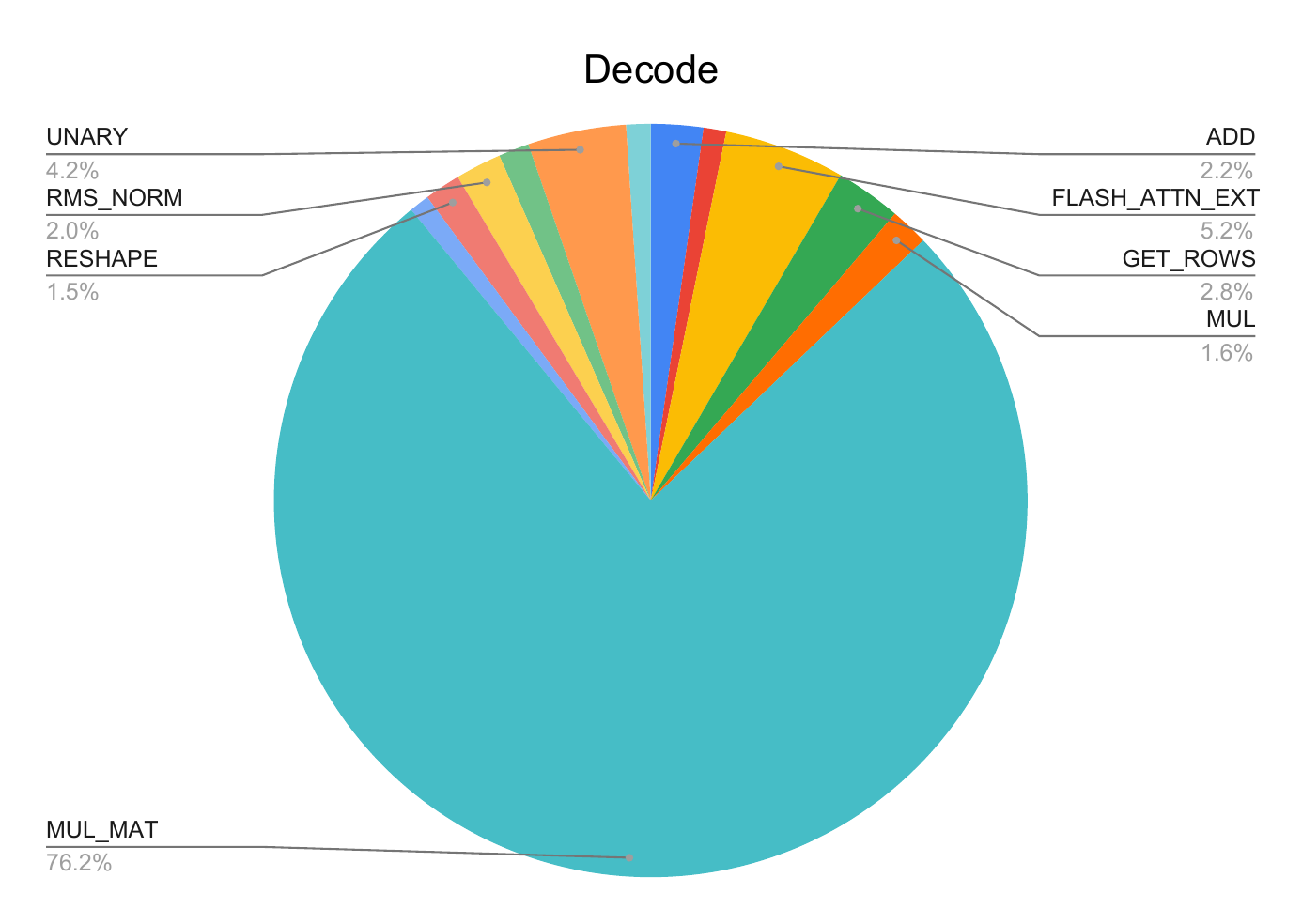}
        \caption{OPs performance in decode stage}
        \label{fig:decode_ops}
    \end{subfigure}
    \caption{Profiling result of different OPs in prefill and decode stages.}
    \label{fig:performance_comparison_of_all_ops}
\end{figure}

\begin{figure}[htp]
    \centering
    \begin{subfigure}[b]{0.48\textwidth}
        \centering
        \includegraphics[width=\textwidth]{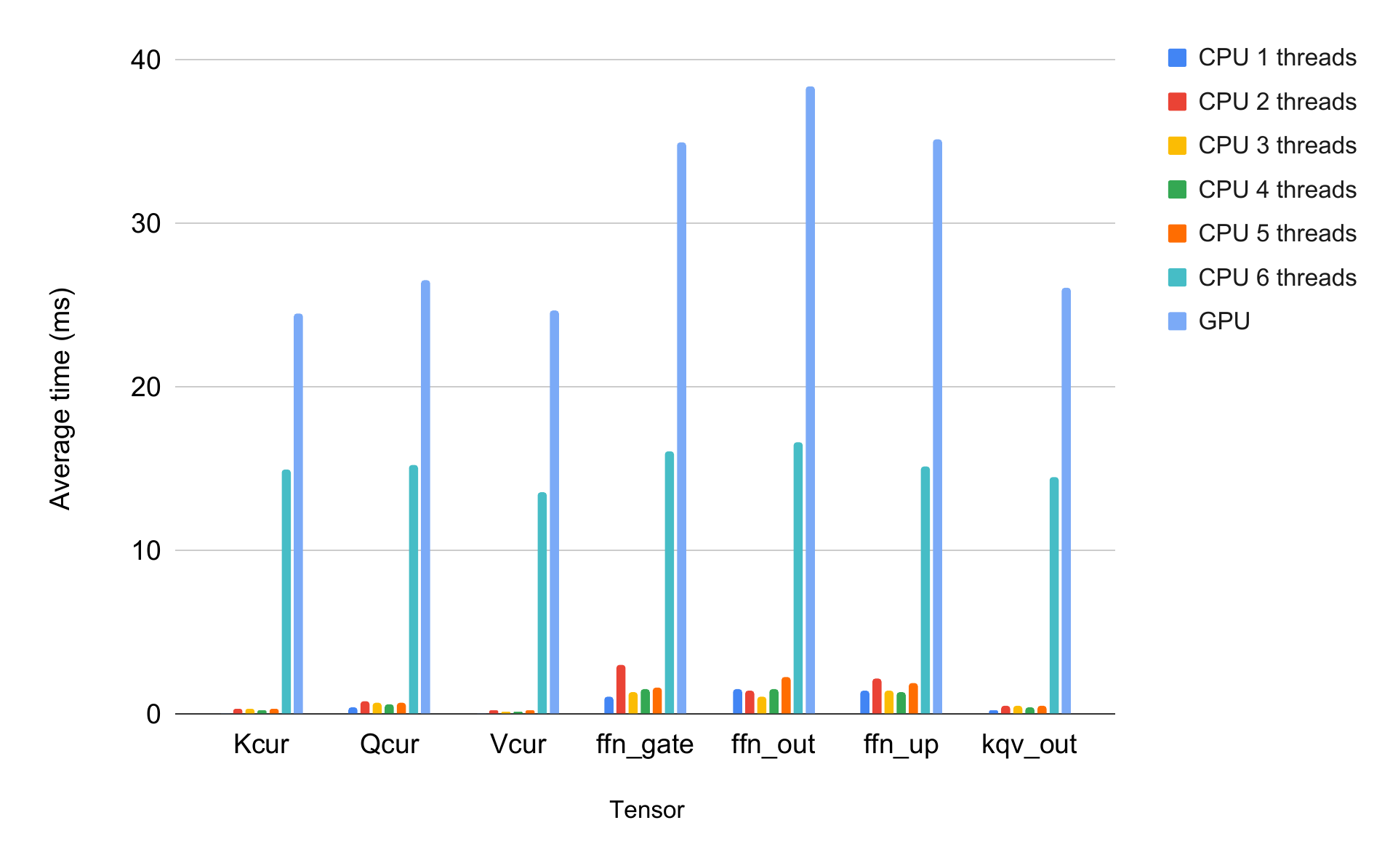}
        \caption{MatMul performance in prefill stage}
        \label{fig:prefill-matmuls}
    \end{subfigure}
    \hfill
    \begin{subfigure}[b]{0.48\textwidth}
        \centering
        \includegraphics[width=\textwidth]{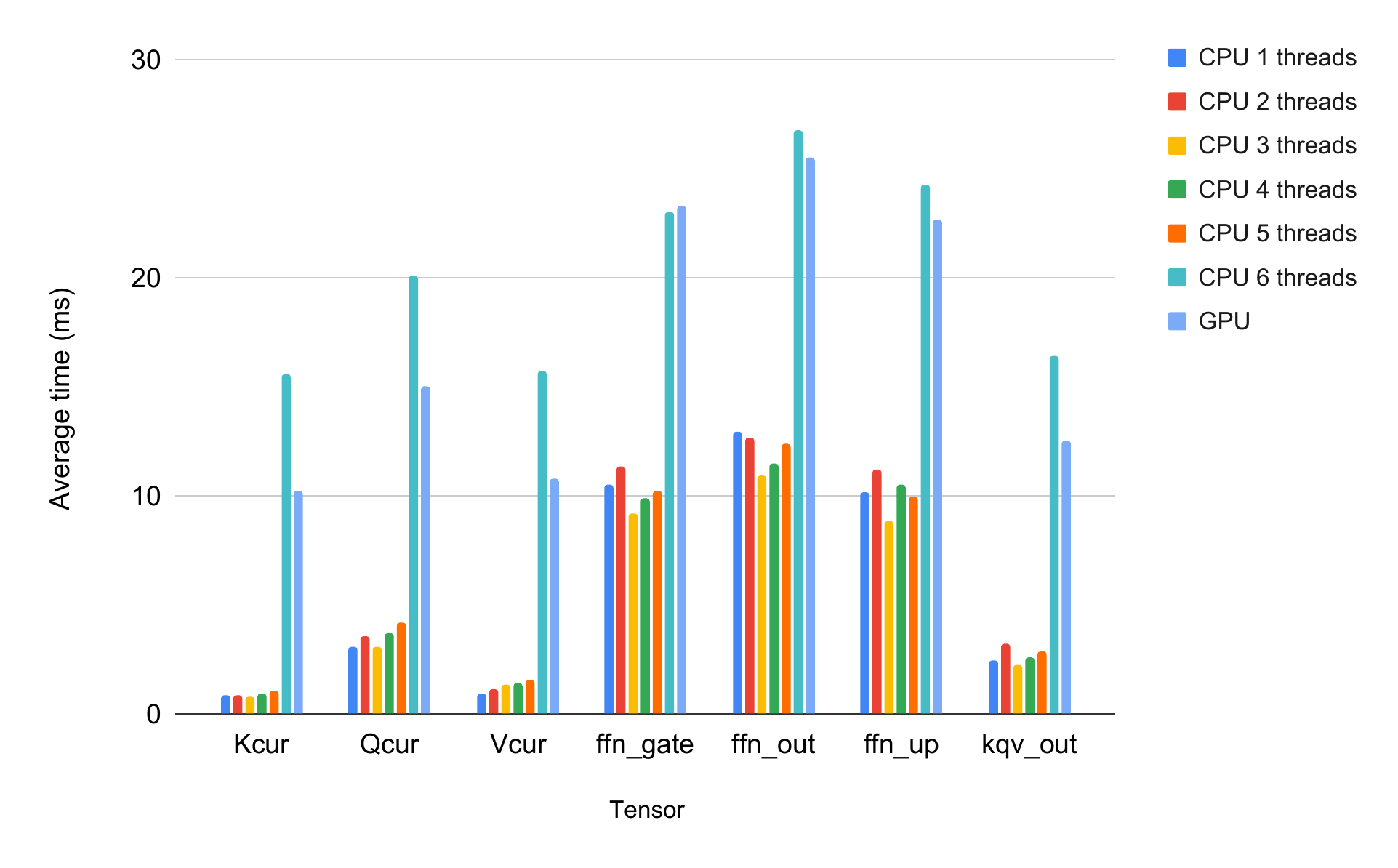}
        \caption{MatMul performance in decode stage}
        \label{fig:decode-matmuls}
    \end{subfigure}
    \caption{MatMul performance comparison between prefill and decode stages.}
    \label{fig:performance_comparison_of_all_mulmats}
\end{figure}

\section{Discussion} \label{Discussion}

\subsection{Profiling Inference Bottlenecks}

To understand why the CPU outperforms the GPU in certain models, we selected the Llama-3.2-1B-F16 model as our target due to its balanced model size and superior performance compared to 0.5B models, along with strong community support. To gain deeper insights into the performance constraints of running Llama-3.2-1B-F16 in a CPU-only environment, we conducted a detailed analysis of execution time distribution across various operations. As shown in Figure~\ref{fig:prefill_ops}, matrix multiplication (\texttt{GGML\_OP\_MUL\_MAT}) dominates the computation during the \textbf{prefill} phase, accounting for \textbf{87.6\%} of the total execution time. A similar trend is observed in the \textbf{decode} phase (Figure~\ref{fig:decode_ops}), where matrix multiplications remain the primary bottleneck, albeit with a slightly reduced share of \textbf{76.2\%}. These findings confirm that General Matrix Multiplication (GEMM) operations are the key computational bottleneck in LLaMA inference.

Given these findings, optimization efforts should focus on accelerating GEMM computations, as they dictate overall inference efficiency. Several avenues merit exploration:
\begin{itemize}
    \item Utilizing ARM SVE SIMD and other architecture-specific instruction sets to accelerate matrix multiplications.
    \item Implementing cache-aware execution strategies to mitigate memory bandwidth constraints.
    \item Exploring hybrid execution models that dynamically partition workloads across CPU and GPU based on compute intensity and memory locality.
\end{itemize}

\subsection{Breakdown of GEMM Operations in LLaMA Inference}

The LLaMA architecture relies on seven distinct matrix multiplications within each decoder layer:
\begin{itemize}
    \item \textbf{Self-Attention Block:} Query (Qcur), Key (Kcur), Value (Vcur), and Attention Output (kqv\_out).
    \item \textbf{Feedforward Network (FFN) Block:} FFN Up (ffn\_up), FFN Gate (ffn\_gate), and FFN Down (ffn\_down).
\end{itemize}

As illustrated in Figures~\ref{fig:prefill-matmuls} and~\ref{fig:decode-matmuls}, profiling results indicate that the FFN block incurs the highest computational cost. This is expected, as the FFN contains two large GEMM operations (\texttt{ffn\_up} and \texttt{ffn\_down}), which operate over a significantly larger hidden dimension than the model's input dimension, amplifying the computational burden.

Despite efforts to improve performance through parallelization—specifically modifying the execution strategy from serial to topological graph execution and distributing workloads concurrently across multiple computational backends—we did not observe significant speedups or improvements in inference throughput. Currently, we cannot precisely pinpoint the factors causing this outcome. The inability to gain deeper insights stems from the limited low-level system access provided by iOS, preventing us from thoroughly profiling execution behaviors, system calls, or resource allocation patterns at runtime.

\section{Topological-Based Graph Execution} \label{Topological-Based Graph Execution}

\subsection{Implementation of Graph-Level Parallelism}
After identifying \texttt{MatMul} as the dominant computational bottleneck (Figures~\ref{fig:prefill_ops} and~\ref{fig:decode_ops}), we explored an alternative execution strategy that leverages compute graph parallelism and \textbf{topological order scheduling}. While llama.cpp’s default execution pipeline processes operations sequentially with low-level tensor parallelism, our approach builds a higher-level compute graph to dispatch independent \texttt{MatMul} operations in parallel.

\begin{figure}[htp]
    \centering
    \includegraphics[width=\columnwidth]{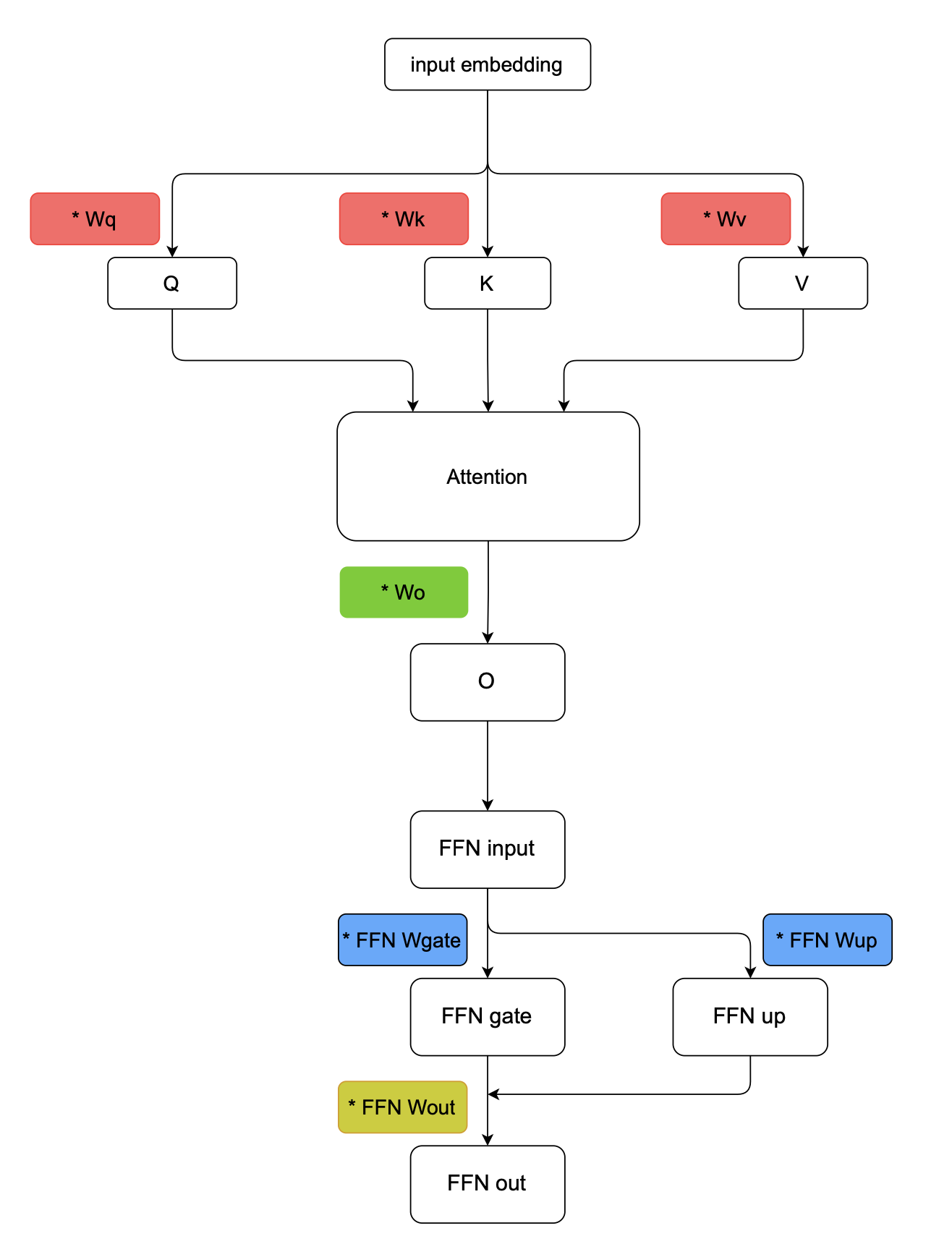}
    \caption{Simplified compute graph of decoder block in Transformer model, matrix multiplication with no dependencies are are colored with same color}
    \label{fig:simplified_cgraph}
\end{figure}

As illustrated in the simplified compute graph (Figure~\ref{fig:simplified_cgraph}), certain nodes—particularly matrix multiplications—exhibit no dependencies on each other, presenting an opportunity for concurrent execution. This independence is especially evident in the self-attention mechanism, where the Q, K, and V \texttt{MatMul} in the attention block, as well as the FFN gate and FFN up \texttt{MatMul} in the FFN block, can be executed in parallel. Unlike tensor parallelism, which partitions a \emph{single} \texttt{MatMul} operation, our approach exploits graph-level parallelism to schedule \emph{multiple} \texttt{MatMul} operations concurrently.

To implement this in llama.cpp, we modified the scheduler to:
\begin{enumerate}
    \item Dynamically analyze the compute graph to detect independent operations.
    \item Schedule independent \texttt{MatMul} operations across available hardware backends (e.g., CPU cores, GPU).
    \item Enforce topological ordering to respect data dependencies while maximizing concurrency.
\end{enumerate}

\subsection{Performance Evaluation of Graph-Level Parallelism}
\begin{figure}[h]
    \centering
    \includegraphics[width=\columnwidth]{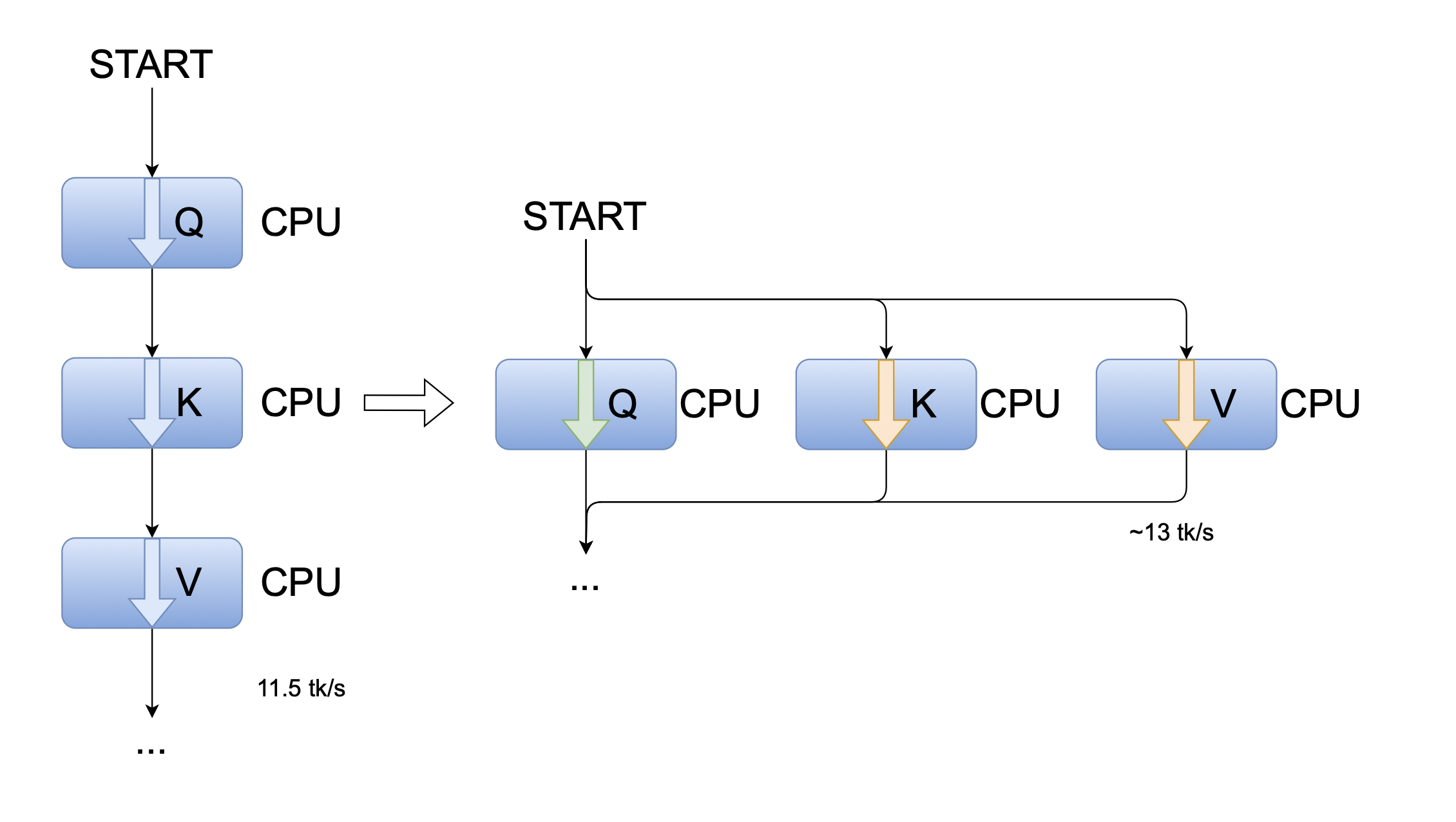}
    \caption{Comparison of the execution flow in llama.cpp before and after introducing graph-level parallelism. The baseline (Left) sequential execution processes Q, K, and V in series, achieving 11.5 tokens per second (tk/s). In contrast, Version 1 (Right) schedules these operations in parallel, increasing throughput to ~13 tk/s.}
    \label{fig:base_2_v1}
\end{figure}

\begin{figure}[h]
    \centering
    \includegraphics[width=\columnwidth]{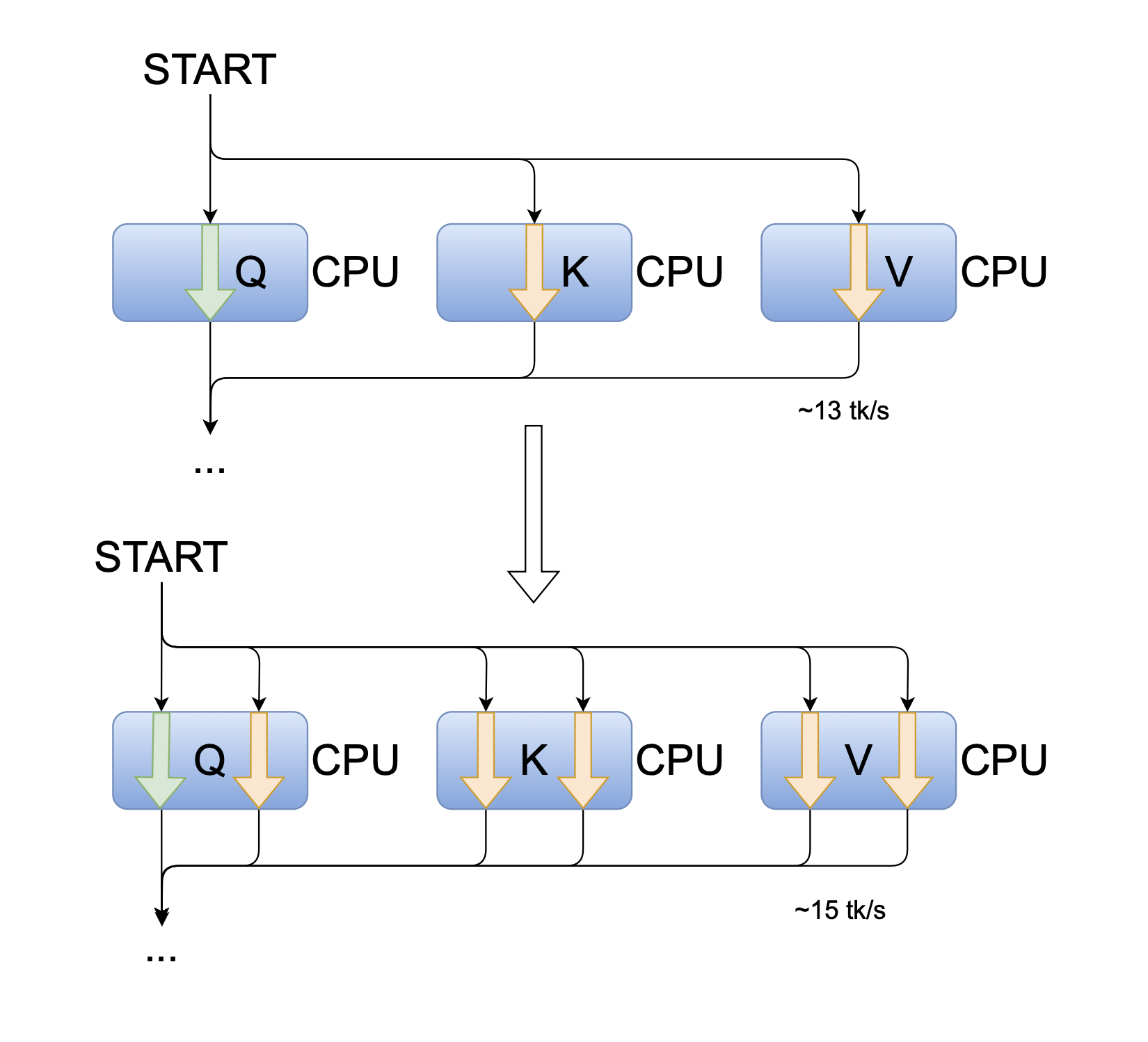}
    \caption{Transition from Version 1 (Top, graph-level parallelism) to Version 2 (Bottom), which incorporates tensor-level parallelism. By leveraging multiple threads per operation, Version 2 achieves further performance improvements, reaching ~15 tk/s.}
    \label{fig:v1_2_v2}
\end{figure}

\begin{figure}[h]
    \centering
    \includegraphics[width=\columnwidth]{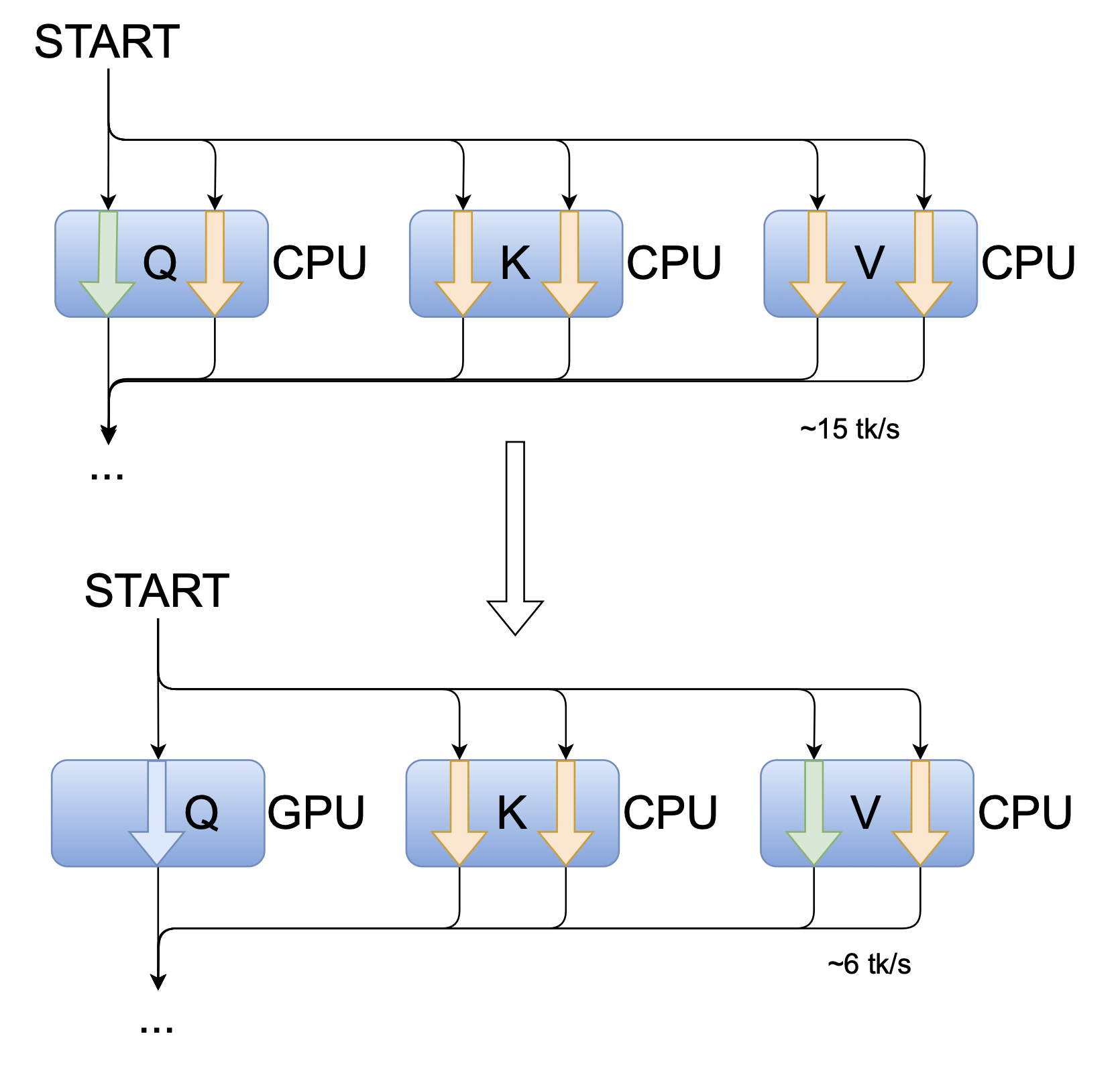}
    \caption{Transition from Version 2 (Top, graph + tensor parallelism) to Version 3 (Bottom), where workloads are distributed across multiple hardware devices (CPU + GPU). Despite expectations of improved performance, Version 3 experiences a significant drop to ~6 tk/s, likely due to memory synchronization overheads between CPU and GPU.}
    \label{fig:v2_2_v3}
\end{figure}

We evaluated our execution model across three successive versions:
\begin{itemize}
    \item \textbf{Version 1:} Introduced graph-level parallelism, improving inference speed to $\sim$13 tokens per second (tk/s), nearly matching default GPU-enabled execution (Figure~\ref{fig:base_2_v1}).
    \item \textbf{Version 2:} Augmented Version~1 with tensor-level parallelism, further increasing throughput to $\sim$15 tk/s (Figure~\ref{fig:v1_2_v2}).
    \item \textbf{Version 3:} Attempted to distribute the graph + tensor workloads \emph{across CPU and GPU}. Instead of improving performance, the throughput dropped significantly to $\sim$6 tk/s (Figure~\ref{fig:v2_2_v3}).
\end{itemize}

\subsection{Analysis of Performance Degradation in Version~3}
Contrary to our expectations, Version~3 exhibited a substantial performance decline when offloading part of the graph to the GPU. We hypothesize several contributing factors:
\begin{itemize}
    \item \textbf{Memory Transfer Overheads:} Even though Apple GPUs use unified memory, runtime allocations and metadata synchronization (e.g., Metal buffers) can incur significant overhead on small-batch operations.
    \item \textbf{Thread Scheduling Conflicts:} Dispatching kernels to the GPU in parallel with CPU threads sometimes introduces scheduling contention, leading to idle time on both CPU and GPU.
    \item \textbf{Underutilized GPU for Small GEMMs:} For lower batch sizes or smaller matrix shapes, GPU kernel launch overhead can overshadow potential gains from hardware acceleration.
\end{itemize}
Hence, while graph-level parallelism was successful on the CPU (Versions~1 and~2), extending it to a heterogeneous environment (CPU + GPU) underscored the complexity of balancing device-specific overheads and concurrency.

\subsection{CPU vs.\ GPU Performance in GEMM Execution}

Despite the inherent parallelism of matrix multiplications, profiling results indicate that CPU-based execution can outperform GPU-based execution in single-batch inference, particularly in small-scale models. The reasons include:
\begin{itemize}
    \item \textbf{Reduced Kernel Launch Overheads:} CPUs can execute small GEMMs more directly, whereas GPUs pay a nontrivial cost for launching kernels on smaller workloads.
    \item \textbf{Efficient Thread Utilization:} Modern CPU architectures with multiple cores (and hyper-threading) handle small, frequent operations effectively, often surpassing GPU throughput on tiny matrices.
\end{itemize}

\subsection{Future Directions for Optimization}

Given that GEMM operations constitute the bulk of inference computation, future work should explore:
\begin{itemize}
    \item \textbf{Hardware-Aware Hybrid Execution:} Dynamically dispatching specific operations (e.g., self-attention vs.\ FFN) to CPU, GPU, or NPU based on workload characteristics.
    \item \textbf{Quantization-Aware Scaling:} Evaluating the impact of lower-bit quantization (e.g., Q2, Q3) on memory-bound execution to reduce CPU inference latency.
    \item \textbf{Fine-Grained Device Partitioning:} Investigating strategies to avoid large overheads in CPU-GPU data transfers, for instance by batching multiple tokens or layers to amortize synchronization costs.
\end{itemize}

Ultimately, a hardware-adaptive approach that tailors parallelization and synchronization strategies to each device’s capabilities is crucial for efficient LLM inference. 

\section{Limitations} \label{Limitations}

While this study provides a detailed performance analysis of Llama 3.2-1B-F16 on mobile hardware, several key challenges remain in fully characterizing computational bottlenecks. These limitations highlight the complexities of profiling and optimizing LLM inference on constrained mobile platforms.

\subsection{Challenges in Function-Level Profiling with Hardware Performance Counters}

Despite successfully collecting hardware performance counter data, accurately isolating performance metrics at the function level proved challenging. The profiling tools available on mobile devices impose constraints that prevent fine-grained attribution of memory bandwidth usage and execution latency to individual GEMM operations. As a result, our analysis cannot precisely quantify how specific layers contribute to overall computational and memory bottlenecks. Addressing this limitation requires either low-overhead instrumentation or deeper integration with vendor-provided profiling APIs.

\subsection{GPU Profiling Instability with Metal Debugger}

Efforts to capture GPU performance counters using Apple’s Metal Debugger were hindered by instability, with repeated application crashes occurring during data transfer between the iPhone and macOS. These failures suggest either an internal buffer limitation or a breakdown in the data transfer protocol when handling large inference traces. Without direct visibility into Metal's internal profiling mechanisms, diagnosing the root cause remains challenging. To mitigate these limitations, future work should explore:
\begin{itemize}
    \item Profiling smaller models (e.g., 0.8B parameters) to determine if data size triggers instability.
    \item Developing custom Metal kernel instrumentation to extract execution statistics outside the debugger.
    \item Investigating Metal's memory hierarchy to assess whether excessive traffic contributes to instability.
\end{itemize}

\subsection{Device-Specific Findings}

Our study focuses on the iPhone 15 Pro, leveraging its A17 Pro 2P+4E CPU and Apple-designed mobile GPU. While these results provide valuable insights, they may not generalize to other ARM-based mobile platforms such as Qualcomm Snapdragon or Google Tensor due to architectural differences in cache hierarchies, memory bandwidth, and GPU compute efficiency. Expanding the study to multiple hardware configurations would help assess the broader applicability of our findings across mobile AI inference pipelines.

\subsection{Limited Scope of Model Size and Precision}

This work primarily evaluates a 1B parameter model using F16 and Q4 precision, providing a representative but narrow perspective on LLM inference efficiency. Larger models (e.g., 3B, 7B) introduce distinct scaling effects that could shift CPU-GPU performance trade-offs. Additionally, ultra-low-bit quantization (e.g., Q2, Q3) remains unexplored, despite its potential to alleviate memory bandwidth constraints and improve CPU efficiency. A more comprehensive study would assess the impact of varying model scales and precision levels on mobile inference.

\subsection{Power and Thermal Constraints}

Although the experiments were conducted in a controlled, ice-cooled environment to mitigate thermal throttling, this does not reflect real-world deployment conditions, where sustained workloads induce heat buildup and frequency scaling. Furthermore, we did not measure power consumption, limiting our ability to assess energy efficiency across CPU and GPU execution paths. Future research should incorporate real-world thermal profiling and power analysis to evaluate sustained performance under typical operating scenarios.

\section{Conclusion} \label{Conclusion}

Large language models (LLMs) are increasingly deployed on mobile devices, yet the optimal execution strategy remains an open question. This study systematically evaluates the inference performance of Llama 3.2-1B across different execution backends, comparing CPU-only computation to GPU-accelerated execution under various precision settings. Contrary to the prevailing assumption that GPUs always provide superior performance, our findings reveal a more nuanced picture:

\begin{itemize}
    \item Under carefully tuned configurations, CPU execution can surpass GPU performance, particularly when leveraging an optimal number of threads.
    \item The inference workload is heavily dominated by GEMM operations in the feedforward network (FFN), making them the primary bottleneck.
    \item Lower-bit quantization (Q4) provides substantial speedups, emerging as a practical choice for applications where efficiency is prioritized over precision.
    \item GPUs face inefficiencies stemming from kernel launch overhead, memory bandwidth constraints, and the challenge of efficiently partitioning workloads for smaller models.
\end{itemize}

These observations suggest that achieving optimal LLM inference performance on mobile devices requires a hardware-aware execution strategy that effectively balances computation across available resources. To this end, future work should explore hybrid execution strategies that dynamically allocate workloads between CPU and GPU based on runtime conditions. Further optimizations in GEMM computations, particularly through ARM-specific instruction tuning, could provide significant gains. Additionally, the role of quantization in reducing memory footprint and improving efficiency warrants deeper investigation.

Beyond computational optimizations, our findings also highlight a critical gap in current profiling tools for mobile AI workloads. Limitations in function-level performance attribution and GPU execution tracing hinder deeper analysis and optimization. Addressing these gaps through improved profiling methodologies will be essential for driving further advancements in on-device LLM performance.

Overall, our study underscores the need for a systematic evaluation of CPU-GPU trade-offs in mobile LLM inference, paving the way for more efficient deployment strategies in resource-constrained environments.

\printbibliography

@misc{touvron2023llamaopenefficientfoundation,
      title={LLaMA: Open and Efficient Foundation Language Models}, 
      author={Hugo Touvron and Thibaut Lavril and Gautier Izacard and Xavier Martinet and Marie-Anne Lachaux and Timothée Lacroix and Baptiste Rozière and Naman Goyal and Eric Hambro and Faisal Azhar and Aurelien Rodriguez and Armand Joulin and Edouard Grave and Guillaume Lample},
      year={2023},
      eprint={2302.13971},
      archivePrefix={arXiv},
      primaryClass={cs.CL},
      url={https://arxiv.org/abs/2302.13971}, 
}

@misc{frantar2023gptqaccurateposttrainingquantization,
      title={GPTQ: Accurate Post-Training Quantization for Generative Pre-trained Transformers}, 
      author={Elias Frantar and Saleh Ashkboos and Torsten Hoefler and Dan Alistarh},
      year={2023},
      eprint={2210.17323},
      archivePrefix={arXiv},
      primaryClass={cs.LG},
      url={https://arxiv.org/abs/2210.17323}, 
}

@misc{llama.cpp,
  author = {Georgi Gerganov},
  title = {llama.cpp},
  year = {2023},
  publisher = {GitHub},
  journal = {GitHub repository},
  howpublished = {\url{https://github.com/ggerganov/llama.cpp}},
}

@misc{jiang2020mnnuniversalefficientinference,
      title={MNN: A Universal and Efficient Inference Engine}, 
      author={Xiaotang Jiang and Huan Wang and Yiliu Chen and Ziqi Wu and Lichuan Wang and Bin Zou and Yafeng Yang and Zongyang Cui and Yu Cai and Tianhang Yu and Chengfei Lv and Zhihua Wu},
      year={2020},
      eprint={2002.12418},
      archivePrefix={arXiv},
      primaryClass={cs.CV},
      url={https://arxiv.org/abs/2002.12418}, 
}

@misc{song2024powerinferfastlargelanguage,
      title={PowerInfer: Fast Large Language Model Serving with a Consumer-grade GPU}, 
      author={Yixin Song and Zeyu Mi and Haotong Xie and Haibo Chen},
      year={2024},
      eprint={2312.12456},
      archivePrefix={arXiv},
      primaryClass={cs.LG},
      url={https://arxiv.org/abs/2312.12456}, 
}

@misc{executorch,
  author = {PyTorch},
  title = {ExecuTorch},
  year = {2024},
  publisher = {GitHub},
  journal = {GitHub repository},
  howpublished = {\url{https://github.com/pytorch/executorch}},
}

@misc{lugaresi2019mediapipeframeworkbuildingperception,
      title={MediaPipe: A Framework for Building Perception Pipelines}, 
      author={Camillo Lugaresi and Jiuqiang Tang and Hadon Nash and Chris McClanahan and Esha Uboweja and Michael Hays and Fan Zhang and Chuo-Ling Chang and Ming Guang Yong and Juhyun Lee and Wan-Teh Chang and Wei Hua and Manfred Georg and Matthias Grundmann},
      year={2019},
      eprint={1906.08172},
      archivePrefix={arXiv},
      primaryClass={cs.DC},
      url={https://arxiv.org/abs/1906.08172}, 
}

@misc{grattafiori2024llama3herdmodels,
      title={The Llama 3 Herd of Models}, 
      author={Aaron Grattafiori and Abhimanyu Dubey and Abhinav Jauhri and Abhinav Pandey and Abhishek Kadian and Ahmad Al-Dahle and Aiesha Letman and Akhil Mathur and Alan Schelten and Alex Vaughan and Amy Yang and Angela Fan and Anirudh Goyal and Anthony Hartshorn and Aobo Yang and Archi Mitra and Archie Sravankumar and Artem Korenev and Arthur Hinsvark and Arun Rao and Aston Zhang and Aurelien Rodriguez and Austen Gregerson and Ava Spataru and Baptiste Roziere and Bethany Biron and Binh Tang and Bobbie Chern and Charlotte Caucheteux and Chaya Nayak and Chloe Bi and Chris Marra and Chris McConnell and Christian Keller and Christophe Touret and Chunyang Wu and Corinne Wong and Cristian Canton Ferrer and Cyrus Nikolaidis and Damien Allonsius and Daniel Song and Danielle Pintz and Danny Livshits and Danny Wyatt and David Esiobu and Dhruv Choudhary and Dhruv Mahajan and Diego Garcia-Olano and Diego Perino and Dieuwke Hupkes and Egor Lakomkin and Ehab AlBadawy and Elina Lobanova and Emily Dinan and Eric Michael Smith and Filip Radenovic and Francisco Guzmán and Frank Zhang and Gabriel Synnaeve and Gabrielle Lee and Georgia Lewis Anderson and Govind Thattai and Graeme Nail and Gregoire Mialon and Guan Pang and Guillem Cucurell and Hailey Nguyen and Hannah Korevaar and Hu Xu and Hugo Touvron and Iliyan Zarov and Imanol Arrieta Ibarra and Isabel Kloumann and Ishan Misra and Ivan Evtimov and Jack Zhang and Jade Copet and Jaewon Lee and Jan Geffert and Jana Vranes and Jason Park and Jay Mahadeokar and Jeet Shah and Jelmer van der Linde and Jennifer Billock and Jenny Hong and Jenya Lee and Jeremy Fu and Jianfeng Chi and Jianyu Huang and Jiawen Liu and Jie Wang and Jiecao Yu and Joanna Bitton and Joe Spisak and Jongsoo Park and Joseph Rocca and Joshua Johnstun and Joshua Saxe and Junteng Jia and Kalyan Vasuden Alwala and Karthik Prasad and Kartikeya Upasani and Kate Plawiak and Ke Li and Kenneth Heafield and Kevin Stone and Khalid El-Arini and Krithika Iyer and Kshitiz Malik and Kuenley Chiu and Kunal Bhalla and Kushal Lakhotia and Lauren Rantala-Yeary and Laurens van der Maaten and Lawrence Chen and Liang Tan and Liz Jenkins and Louis Martin and Lovish Madaan and Lubo Malo and Lukas Blecher and Lukas Landzaat and Luke de Oliveira and Madeline Muzzi and Mahesh Pasupuleti and Mannat Singh and Manohar Paluri and Marcin Kardas and Maria Tsimpoukelli and Mathew Oldham and Mathieu Rita and Maya Pavlova and Melanie Kambadur and Mike Lewis and Min Si and Mitesh Kumar Singh and Mona Hassan and Naman Goyal and Narjes Torabi and Nikolay Bashlykov and Nikolay Bogoychev and Niladri Chatterji and Ning Zhang and Olivier Duchenne and Onur Çelebi and Patrick Alrassy and Pengchuan Zhang and Pengwei Li and Petar Vasic and Peter Weng and Prajjwal Bhargava and Pratik Dubal and Praveen Krishnan and Punit Singh Koura and Puxin Xu and Qing He and Qingxiao Dong and Ragavan Srinivasan and Raj Ganapathy and Ramon Calderer and Ricardo Silveira Cabral and Robert Stojnic and Roberta Raileanu and Rohan Maheswari and Rohit Girdhar and Rohit Patel and Romain Sauvestre and Ronnie Polidoro and Roshan Sumbaly and Ross Taylor and Ruan Silva and Rui Hou and Rui Wang and Saghar Hosseini and Sahana Chennabasappa and Sanjay Singh and Sean Bell and Seohyun Sonia Kim and Sergey Edunov and Shaoliang Nie and Sharan Narang and Sharath Raparthy and Sheng Shen and Shengye Wan and Shruti Bhosale and Shun Zhang and Simon Vandenhende and Soumya Batra and Spencer Whitman and Sten Sootla and Stephane Collot and Suchin Gururangan and Sydney Borodinsky and Tamar Herman and Tara Fowler and Tarek Sheasha and Thomas Georgiou and Thomas Scialom and Tobias Speckbacher and Todor Mihaylov and Tong Xiao and Ujjwal Karn and Vedanuj Goswami and Vibhor Gupta and Vignesh Ramanathan and Viktor Kerkez and Vincent Gonguet and Virginie Do and Vish Vogeti and Vítor Albiero and Vladan Petrovic and Weiwei Chu and Wenhan Xiong and Wenyin Fu and Whitney Meers and Xavier Martinet and Xiaodong Wang and Xiaofang Wang and Xiaoqing Ellen Tan and Xide Xia and Xinfeng Xie and Xuchao Jia and Xuewei Wang and Yaelle Goldschlag and Yashesh Gaur and Yasmine Babaei and Yi Wen and Yiwen Song and Yuchen Zhang and Yue Li and Yuning Mao and Zacharie Delpierre Coudert and Zheng Yan and Zhengxing Chen and Zoe Papakipos and Aaditya Singh and Aayushi Srivastava and Abha Jain and Adam Kelsey and Adam Shajnfeld and Adithya Gangidi and Adolfo Victoria and Ahuva Goldstand and Ajay Menon and Ajay Sharma and Alex Boesenberg and Alexei Baevski and Allie Feinstein and Amanda Kallet and Amit Sangani and Amos Teo and Anam Yunus and Andrei Lupu and Andres Alvarado and Andrew Caples and Andrew Gu and Andrew Ho and Andrew Poulton and Andrew Ryan and Ankit Ramchandani and Annie Dong and Annie Franco and Anuj Goyal and Aparajita Saraf and Arkabandhu Chowdhury and Ashley Gabriel and Ashwin Bharambe and Assaf Eisenman and Azadeh Yazdan and Beau James and Ben Maurer and Benjamin Leonhardi and Bernie Huang and Beth Loyd and Beto De Paola and Bhargavi Paranjape and Bing Liu and Bo Wu and Boyu Ni and Braden Hancock and Bram Wasti and Brandon Spence and Brani Stojkovic and Brian Gamido and Britt Montalvo and Carl Parker and Carly Burton and Catalina Mejia and Ce Liu and Changhan Wang and Changkyu Kim and Chao Zhou and Chester Hu and Ching-Hsiang Chu and Chris Cai and Chris Tindal and Christoph Feichtenhofer and Cynthia Gao and Damon Civin and Dana Beaty and Daniel Kreymer and Daniel Li and David Adkins and David Xu and Davide Testuggine and Delia David and Devi Parikh and Diana Liskovich and Didem Foss and Dingkang Wang and Duc Le and Dustin Holland and Edward Dowling and Eissa Jamil and Elaine Montgomery and Eleonora Presani and Emily Hahn and Emily Wood and Eric-Tuan Le and Erik Brinkman and Esteban Arcaute and Evan Dunbar and Evan Smothers and Fei Sun and Felix Kreuk and Feng Tian and Filippos Kokkinos and Firat Ozgenel and Francesco Caggioni and Frank Kanayet and Frank Seide and Gabriela Medina Florez and Gabriella Schwarz and Gada Badeer and Georgia Swee and Gil Halpern and Grant Herman and Grigory Sizov and Guangyi and Zhang and Guna Lakshminarayanan and Hakan Inan and Hamid Shojanazeri and Han Zou and Hannah Wang and Hanwen Zha and Haroun Habeeb and Harrison Rudolph and Helen Suk and Henry Aspegren and Hunter Goldman and Hongyuan Zhan and Ibrahim Damlaj and Igor Molybog and Igor Tufanov and Ilias Leontiadis and Irina-Elena Veliche and Itai Gat and Jake Weissman and James Geboski and James Kohli and Janice Lam and Japhet Asher and Jean-Baptiste Gaya and Jeff Marcus and Jeff Tang and Jennifer Chan and Jenny Zhen and Jeremy Reizenstein and Jeremy Teboul and Jessica Zhong and Jian Jin and Jingyi Yang and Joe Cummings and Jon Carvill and Jon Shepard and Jonathan McPhie and Jonathan Torres and Josh Ginsburg and Junjie Wang and Kai Wu and Kam Hou U and Karan Saxena and Kartikay Khandelwal and Katayoun Zand and Kathy Matosich and Kaushik Veeraraghavan and Kelly Michelena and Keqian Li and Kiran Jagadeesh and Kun Huang and Kunal Chawla and Kyle Huang and Lailin Chen and Lakshya Garg and Lavender A and Leandro Silva and Lee Bell and Lei Zhang and Liangpeng Guo and Licheng Yu and Liron Moshkovich and Luca Wehrstedt and Madian Khabsa and Manav Avalani and Manish Bhatt and Martynas Mankus and Matan Hasson and Matthew Lennie and Matthias Reso and Maxim Groshev and Maxim Naumov and Maya Lathi and Meghan Keneally and Miao Liu and Michael L. Seltzer and Michal Valko and Michelle Restrepo and Mihir Patel and Mik Vyatskov and Mikayel Samvelyan and Mike Clark and Mike Macey and Mike Wang and Miquel Jubert Hermoso and Mo Metanat and Mohammad Rastegari and Munish Bansal and Nandhini Santhanam and Natascha Parks and Natasha White and Navyata Bawa and Nayan Singhal and Nick Egebo and Nicolas Usunier and Nikhil Mehta and Nikolay Pavlovich Laptev and Ning Dong and Norman Cheng and Oleg Chernoguz and Olivia Hart and Omkar Salpekar and Ozlem Kalinli and Parkin Kent and Parth Parekh and Paul Saab and Pavan Balaji and Pedro Rittner and Philip Bontrager and Pierre Roux and Piotr Dollar and Polina Zvyagina and Prashant Ratanchandani and Pritish Yuvraj and Qian Liang and Rachad Alao and Rachel Rodriguez and Rafi Ayub and Raghotham Murthy and Raghu Nayani and Rahul Mitra and Rangaprabhu Parthasarathy and Raymond Li and Rebekkah Hogan and Robin Battey and Rocky Wang and Russ Howes and Ruty Rinott and Sachin Mehta and Sachin Siby and Sai Jayesh Bondu and Samyak Datta and Sara Chugh and Sara Hunt and Sargun Dhillon and Sasha Sidorov and Satadru Pan and Saurabh Mahajan and Saurabh Verma and Seiji Yamamoto and Sharadh Ramaswamy and Shaun Lindsay and Shaun Lindsay and Sheng Feng and Shenghao Lin and Shengxin Cindy Zha and Shishir Patil and Shiva Shankar and Shuqiang Zhang and Shuqiang Zhang and Sinong Wang and Sneha Agarwal and Soji Sajuyigbe and Soumith Chintala and Stephanie Max and Stephen Chen and Steve Kehoe and Steve Satterfield and Sudarshan Govindaprasad and Sumit Gupta and Summer Deng and Sungmin Cho and Sunny Virk and Suraj Subramanian and Sy Choudhury and Sydney Goldman and Tal Remez and Tamar Glaser and Tamara Best and Thilo Koehler and Thomas Robinson and Tianhe Li and Tianjun Zhang and Tim Matthews and Timothy Chou and Tzook Shaked and Varun Vontimitta and Victoria Ajayi and Victoria Montanez and Vijai Mohan and Vinay Satish Kumar and Vishal Mangla and Vlad Ionescu and Vlad Poenaru and Vlad Tiberiu Mihailescu and Vladimir Ivanov and Wei Li and Wenchen Wang and Wenwen Jiang and Wes Bouaziz and Will Constable and Xiaocheng Tang and Xiaojian Wu and Xiaolan Wang and Xilun Wu and Xinbo Gao and Yaniv Kleinman and Yanjun Chen and Ye Hu and Ye Jia and Ye Qi and Yenda Li and Yilin Zhang and Ying Zhang and Yossi Adi and Youngjin Nam and Yu and Wang and Yu Zhao and Yuchen Hao and Yundi Qian and Yunlu Li and Yuzi He and Zach Rait and Zachary DeVito and Zef Rosnbrick and Zhaoduo Wen and Zhenyu Yang and Zhiwei Zhao and Zhiyu Ma},
      year={2024},
      eprint={2407.21783},
      archivePrefix={arXiv},
      primaryClass={cs.AI},
      url={https://arxiv.org/abs/2407.21783}, 
}

@misc{yang2024qwen2technicalreport,
      title={Qwen2 Technical Report}, 
      author={An Yang and Baosong Yang and Binyuan Hui and Bo Zheng and Bowen Yu and Chang Zhou and Chengpeng Li and Chengyuan Li and Dayiheng Liu and Fei Huang and Guanting Dong and Haoran Wei and Huan Lin and Jialong Tang and Jialin Wang and Jian Yang and Jianhong Tu and Jianwei Zhang and Jianxin Ma and Jianxin Yang and Jin Xu and Jingren Zhou and Jinze Bai and Jinzheng He and Junyang Lin and Kai Dang and Keming Lu and Keqin Chen and Kexin Yang and Mei Li and Mingfeng Xue and Na Ni and Pei Zhang and Peng Wang and Ru Peng and Rui Men and Ruize Gao and Runji Lin and Shijie Wang and Shuai Bai and Sinan Tan and Tianhang Zhu and Tianhao Li and Tianyu Liu and Wenbin Ge and Xiaodong Deng and Xiaohuan Zhou and Xingzhang Ren and Xinyu Zhang and Xipin Wei and Xuancheng Ren and Xuejing Liu and Yang Fan and Yang Yao and Yichang Zhang and Yu Wan and Yunfei Chu and Yuqiong Liu and Zeyu Cui and Zhenru Zhang and Zhifang Guo and Zhihao Fan},
      year={2024},
      eprint={2407.10671},
      archivePrefix={arXiv},
      primaryClass={cs.CL},
      url={https://arxiv.org/abs/2407.10671}, 
}

@misc{jiang2023mistral7b,
      title={Mistral 7B}, 
      author={Albert Q. Jiang and Alexandre Sablayrolles and Arthur Mensch and Chris Bamford and Devendra Singh Chaplot and Diego de las Casas and Florian Bressand and Gianna Lengyel and Guillaume Lample and Lucile Saulnier and Lélio Renard Lavaud and Marie-Anne Lachaux and Pierre Stock and Teven Le Scao and Thibaut Lavril and Thomas Wang and Timothée Lacroix and William El Sayed},
      year={2023},
      eprint={2310.06825},
      archivePrefix={arXiv},
      primaryClass={cs.CL},
      url={https://arxiv.org/abs/2310.06825}, 
}

@misc{Czerski_2025, title={Implementing small language models (slms) with RAG on embedded devices leading to cost reduction, data privacy, and offline use}, url={https://deepsense.ai/blog/implementing-small-language-models-slms-with-rag-on-embedded-devices-leading-to-cost-reduction-data-privacy-and-offline-use}, journal={deepsense.ai}, author={Czerski, Kamil}, year={2025}, month={Jan}}

@misc{zhang2024tinyllamaopensourcesmalllanguage,
      title={TinyLlama: An Open-Source Small Language Model}, 
      author={Peiyuan Zhang and Guangtao Zeng and Tianduo Wang and Wei Lu},
      year={2024},
      eprint={2401.02385},
      archivePrefix={arXiv},
      primaryClass={cs.CL},
      url={https://arxiv.org/abs/2401.02385}, 
}

@article{javaheripi2023phi,
  title={Phi-2: The surprising power of small language models},
  author={Javaheripi, Mojan and Bubeck, S{\'e}bastien and Abdin, Marah and Aneja, Jyoti and Bubeck, Sebastien and Mendes, Caio C{\'e}sar Teodoro and Chen, Weizhu and Del Giorno, Allie and Eldan, Ronen and Gopi, Sivakanth and others},
  journal={Microsoft Research Blog},
  volume={1},
  number={3},
  pages={3},
  year={2023}
}

@article{abdin2024phi,
  title={Phi-3 technical report: A highly capable language model locally on your phone},
  author={Abdin, Marah and Aneja, Jyoti and Awadalla, Hany and Awadallah, Ahmed and Awan, Ammar Ahmad and Bach, Nguyen and Bahree, Amit and Bakhtiari, Arash and Bao, Jianmin and Behl, Harkirat and others},
  journal={arXiv preprint arXiv:2404.14219},
  year={2024}
}

@article{liu2024mobilellm,
  title={Mobilellm: Optimizing sub-billion parameter language models for on-device use cases},
  author={Liu, Zechun and Zhao, Changsheng and Iandola, Forrest and Lai, Chen and Tian, Yuandong and Fedorov, Igor and Xiong, Yunyang and Chang, Ernie and Shi, Yangyang and Krishnamoorthi, Raghuraman and others},
  journal={arXiv preprint arXiv:2402.14905},
  year={2024}
}

@misc{pham2024slimlmefficientsmalllanguage,
      title={SlimLM: An Efficient Small Language Model for On-Device Document Assistance}, 
      author={Thang M. Pham and Phat T. Nguyen and Seunghyun Yoon and Viet Dac Lai and Franck Dernoncourt and Trung Bui},
      year={2024},
      eprint={2411.09944},
      archivePrefix={arXiv},
      primaryClass={cs.CL},
      url={https://arxiv.org/abs/2411.09944}, 
}

@misc{li2024transformerlitehighefficiencydeploymentlarge,
      title={Transformer-Lite: High-efficiency Deployment of Large Language Models on Mobile Phone GPUs}, 
      author={Luchang Li and Sheng Qian and Jie Lu and Lunxi Yuan and Rui Wang and Qin Xie},
      year={2024},
      eprint={2403.20041},
      archivePrefix={arXiv},
      primaryClass={cs.CL},
      url={https://arxiv.org/abs/2403.20041}, 
}

@misc{appleintelligence, title={Apple Intelligence for developers}, url={https://developer.apple.com/apple-intelligence/}, journal={Apple Developer}, author={Inc., Apple}}

@misc{gunter2024appleintelligencefoundationlanguage,
      title={Apple Intelligence Foundation Language Models}, 
      author={Tom Gunter and Zirui Wang and Chong Wang and Ruoming Pang and Andy Narayanan and Aonan Zhang and Bowen Zhang and Chen Chen and Chung-Cheng Chiu and David Qiu and Deepak Gopinath and Dian Ang Yap and Dong Yin and Feng Nan and Floris Weers and Guoli Yin and Haoshuo Huang and Jianyu Wang and Jiarui Lu and John Peebles and Ke Ye and Mark Lee and Nan Du and Qibin Chen and Quentin Keunebroek and Sam Wiseman and Syd Evans and Tao Lei and Vivek Rathod and Xiang Kong and Xianzhi Du and Yanghao Li and Yongqiang Wang and Yuan Gao and Zaid Ahmed and Zhaoyang Xu and Zhiyun Lu and Al Rashid and Albin Madappally Jose and Alec Doane and Alfredo Bencomo and Allison Vanderby and Andrew Hansen and Ankur Jain and Anupama Mann Anupama and Areeba Kamal and Bugu Wu and Carolina Brum and Charlie Maalouf and Chinguun Erdenebileg and Chris Dulhanty and Dominik Moritz and Doug Kang and Eduardo Jimenez and Evan Ladd and Fangping Shi and Felix Bai and Frank Chu and Fred Hohman and Hadas Kotek and Hannah Gillis Coleman and Jane Li and Jeffrey Bigham and Jeffery Cao and Jeff Lai and Jessica Cheung and Jiulong Shan and Joe Zhou and John Li and Jun Qin and Karanjeet Singh and Karla Vega and Kelvin Zou and Laura Heckman and Lauren Gardiner and Margit Bowler and Maria Cordell and Meng Cao and Nicole Hay and Nilesh Shahdadpuri and Otto Godwin and Pranay Dighe and Pushyami Rachapudi and Ramsey Tantawi and Roman Frigg and Sam Davarnia and Sanskruti Shah and Saptarshi Guha and Sasha Sirovica and Shen Ma and Shuang Ma and Simon Wang and Sulgi Kim and Suma Jayaram and Vaishaal Shankar and Varsha Paidi and Vivek Kumar and Xin Wang and Xin Zheng and Walker Cheng and Yael Shrager and Yang Ye and Yasu Tanaka and Yihao Guo and Yunsong Meng and Zhao Tang Luo and Zhi Ouyang and Alp Aygar and Alvin Wan and Andrew Walkingshaw and Andy Narayanan and Antonie Lin and Arsalan Farooq and Brent Ramerth and Colorado Reed and Chris Bartels and Chris Chaney and David Riazati and Eric Liang Yang and Erin Feldman and Gabriel Hochstrasser and Guillaume Seguin and Irina Belousova and Joris Pelemans and Karen Yang and Keivan Alizadeh Vahid and Liangliang Cao and Mahyar Najibi and Marco Zuliani and Max Horton and Minsik Cho and Nikhil Bhendawade and Patrick Dong and Piotr Maj and Pulkit Agrawal and Qi Shan and Qichen Fu and Regan Poston and Sam Xu and Shuangning Liu and Sushma Rao and Tashweena Heeramun and Thomas Merth and Uday Rayala and Victor Cui and Vivek Rangarajan Sridhar and Wencong Zhang and Wenqi Zhang and Wentao Wu and Xingyu Zhou and Xinwen Liu and Yang Zhao and Yin Xia and Zhile Ren and Zhongzheng Ren},
      year={2024},
      eprint={2407.21075},
      archivePrefix={arXiv},
      primaryClass={cs.AI},
      url={https://arxiv.org/abs/2407.21075}, 
}

@misc{geminiteam2024geminifamilyhighlycapable,
      title={Gemini: A Family of Highly Capable Multimodal Models}, 
      author={Gemini Team and Rohan Anil and Sebastian Borgeaud and Jean-Baptiste Alayrac and Jiahui Yu and Radu Soricut and Johan Schalkwyk and Andrew M. Dai and Anja Hauth and Katie Millican and David Silver and Melvin Johnson and Ioannis Antonoglou and Julian Schrittwieser and Amelia Glaese and Jilin Chen and Emily Pitler and Timothy Lillicrap and Angeliki Lazaridou and Orhan Firat and James Molloy and Michael Isard and Paul R. Barham and Tom Hennigan and Benjamin Lee and Fabio Viola and Malcolm Reynolds and Yuanzhong Xu and Ryan Doherty and Eli Collins and Clemens Meyer and Eliza Rutherford and Erica Moreira and Kareem Ayoub and Megha Goel and Jack Krawczyk and Cosmo Du and Ed Chi and Heng-Tze Cheng and Eric Ni and Purvi Shah and Patrick Kane and Betty Chan and Manaal Faruqui and Aliaksei Severyn and Hanzhao Lin and YaGuang Li and Yong Cheng and Abe Ittycheriah and Mahdis Mahdieh and Mia Chen and Pei Sun and Dustin Tran and Sumit Bagri and Balaji Lakshminarayanan and Jeremiah Liu and Andras Orban and Fabian Güra and Hao Zhou and Xinying Song and Aurelien Boffy and Harish Ganapathy and Steven Zheng and HyunJeong Choe and Ágoston Weisz and Tao Zhu and Yifeng Lu and Siddharth Gopal and Jarrod Kahn and Maciej Kula and Jeff Pitman and Rushin Shah and Emanuel Taropa and Majd Al Merey and Martin Baeuml and Zhifeng Chen and Laurent El Shafey and Yujing Zhang and Olcan Sercinoglu and George Tucker and Enrique Piqueras and Maxim Krikun and Iain Barr and Nikolay Savinov and Ivo Danihelka and Becca Roelofs and Anaïs White and Anders Andreassen and Tamara von Glehn and Lakshman Yagati and Mehran Kazemi and Lucas Gonzalez and Misha Khalman and Jakub Sygnowski and Alexandre Frechette and Charlotte Smith and Laura Culp and Lev Proleev and Yi Luan and Xi Chen and James Lottes and Nathan Schucher and Federico Lebron and Alban Rrustemi and Natalie Clay and Phil Crone and Tomas Kocisky and Jeffrey Zhao and Bartek Perz and Dian Yu and Heidi Howard and Adam Bloniarz and Jack W. Rae and Han Lu and Laurent Sifre and Marcello Maggioni and Fred Alcober and Dan Garrette and Megan Barnes and Shantanu Thakoor and Jacob Austin and Gabriel Barth-Maron and William Wong and Rishabh Joshi and Rahma Chaabouni and Deeni Fatiha and Arun Ahuja and Gaurav Singh Tomar and Evan Senter and Martin Chadwick and Ilya Kornakov and Nithya Attaluri and Iñaki Iturrate and Ruibo Liu and Yunxuan Li and Sarah Cogan and Jeremy Chen and Chao Jia and Chenjie Gu and Qiao Zhang and Jordan Grimstad and Ale Jakse Hartman and Xavier Garcia and Thanumalayan Sankaranarayana Pillai and Jacob Devlin and Michael Laskin and Diego de Las Casas and Dasha Valter and Connie Tao and Lorenzo Blanco and Adrià Puigdomènech Badia and David Reitter and Mianna Chen and Jenny Brennan and Clara Rivera and Sergey Brin and Shariq Iqbal and Gabriela Surita and Jane Labanowski and Abhi Rao and Stephanie Winkler and Emilio Parisotto and Yiming Gu and Kate Olszewska and Ravi Addanki and Antoine Miech and Annie Louis and Denis Teplyashin and Geoff Brown and Elliot Catt and Jan Balaguer and Jackie Xiang and Pidong Wang and Zoe Ashwood and Anton Briukhov and Albert Webson and Sanjay Ganapathy and Smit Sanghavi and Ajay Kannan and Ming-Wei Chang and Axel Stjerngren and Josip Djolonga and Yuting Sun and Ankur Bapna and Matthew Aitchison and Pedram Pejman and Henryk Michalewski and Tianhe Yu and Cindy Wang and Juliette Love and Junwhan Ahn and Dawn Bloxwich and Kehang Han and Peter Humphreys and Thibault Sellam and James Bradbury and Varun Godbole and Sina Samangooei and Bogdan Damoc and Alex Kaskasoli and Sébastien M. R. Arnold and Vijay Vasudevan and Shubham Agrawal and Jason Riesa and Dmitry Lepikhin and Richard Tanburn and Srivatsan Srinivasan and Hyeontaek Lim and Sarah Hodkinson and Pranav Shyam and Johan Ferret and Steven Hand and Ankush Garg and Tom Le Paine and Jian Li and Yujia Li and Minh Giang and Alexander Neitz and Zaheer Abbas and Sarah York and Machel Reid and Elizabeth Cole and Aakanksha Chowdhery and Dipanjan Das and Dominika Rogozińska and Vitaliy Nikolaev and Pablo Sprechmann and Zachary Nado and Lukas Zilka and Flavien Prost and Luheng He and Marianne Monteiro and Gaurav Mishra and Chris Welty and Josh Newlan and Dawei Jia and Miltiadis Allamanis and Clara Huiyi Hu and Raoul de Liedekerke and Justin Gilmer and Carl Saroufim and Shruti Rijhwani and Shaobo Hou and Disha Shrivastava and Anirudh Baddepudi and Alex Goldin and Adnan Ozturel and Albin Cassirer and Yunhan Xu and Daniel Sohn and Devendra Sachan and Reinald Kim Amplayo and Craig Swanson and Dessie Petrova and Shashi Narayan and Arthur Guez and Siddhartha Brahma and Jessica Landon and Miteyan Patel and Ruizhe Zhao and Kevin Villela and Luyu Wang and Wenhao Jia and Matthew Rahtz and Mai Giménez and Legg Yeung and James Keeling and Petko Georgiev and Diana Mincu and Boxi Wu and Salem Haykal and Rachel Saputro and Kiran Vodrahalli and James Qin and Zeynep Cankara and Abhanshu Sharma and Nick Fernando and Will Hawkins and Behnam Neyshabur and Solomon Kim and Adrian Hutter and Priyanka Agrawal and Alex Castro-Ros and George van den Driessche and Tao Wang and Fan Yang and Shuo-yiin Chang and Paul Komarek and Ross McIlroy and Mario Lučić and Guodong Zhang and Wael Farhan and Michael Sharman and Paul Natsev and Paul Michel and Yamini Bansal and Siyuan Qiao and Kris Cao and Siamak Shakeri and Christina Butterfield and Justin Chung and Paul Kishan Rubenstein and Shivani Agrawal and Arthur Mensch and Kedar Soparkar and Karel Lenc and Timothy Chung and Aedan Pope and Loren Maggiore and Jackie Kay and Priya Jhakra and Shibo Wang and Joshua Maynez and Mary Phuong and Taylor Tobin and Andrea Tacchetti and Maja Trebacz and Kevin Robinson and Yash Katariya and Sebastian Riedel and Paige Bailey and Kefan Xiao and Nimesh Ghelani and Lora Aroyo and Ambrose Slone and Neil Houlsby and Xuehan Xiong and Zhen Yang and Elena Gribovskaya and Jonas Adler and Mateo Wirth and Lisa Lee and Music Li and Thais Kagohara and Jay Pavagadhi and Sophie Bridgers and Anna Bortsova and Sanjay Ghemawat and Zafarali Ahmed and Tianqi Liu and Richard Powell and Vijay Bolina and Mariko Iinuma and Polina Zablotskaia and James Besley and Da-Woon Chung and Timothy Dozat and Ramona Comanescu and Xiance Si and Jeremy Greer and Guolong Su and Martin Polacek and Raphaël Lopez Kaufman and Simon Tokumine and Hexiang Hu and Elena Buchatskaya and Yingjie Miao and Mohamed Elhawaty and Aditya Siddhant and Nenad Tomasev and Jinwei Xing and Christina Greer and Helen Miller and Shereen Ashraf and Aurko Roy and Zizhao Zhang and Ada Ma and Angelos Filos and Milos Besta and Rory Blevins and Ted Klimenko and Chih-Kuan Yeh and Soravit Changpinyo and Jiaqi Mu and Oscar Chang and Mantas Pajarskas and Carrie Muir and Vered Cohen and Charline Le Lan and Krishna Haridasan and Amit Marathe and Steven Hansen and Sholto Douglas and Rajkumar Samuel and Mingqiu Wang and Sophia Austin and Chang Lan and Jiepu Jiang and Justin Chiu and Jaime Alonso Lorenzo and Lars Lowe Sjösund and Sébastien Cevey and Zach Gleicher and Thi Avrahami and Anudhyan Boral and Hansa Srinivasan and Vittorio Selo and Rhys May and Konstantinos Aisopos and Léonard Hussenot and Livio Baldini Soares and Kate Baumli and Michael B. Chang and Adrià Recasens and Ben Caine and Alexander Pritzel and Filip Pavetic and Fabio Pardo and Anita Gergely and Justin Frye and Vinay Ramasesh and Dan Horgan and Kartikeya Badola and Nora Kassner and Subhrajit Roy and Ethan Dyer and Víctor Campos Campos and Alex Tomala and Yunhao Tang and Dalia El Badawy and Elspeth White and Basil Mustafa and Oran Lang and Abhishek Jindal and Sharad Vikram and Zhitao Gong and Sergi Caelles and Ross Hemsley and Gregory Thornton and Fangxiaoyu Feng and Wojciech Stokowiec and Ce Zheng and Phoebe Thacker and Çağlar Ünlü and Zhishuai Zhang and Mohammad Saleh and James Svensson and Max Bileschi and Piyush Patil and Ankesh Anand and Roman Ring and Katerina Tsihlas and Arpi Vezer and Marco Selvi and Toby Shevlane and Mikel Rodriguez and Tom Kwiatkowski and Samira Daruki and Keran Rong and Allan Dafoe and Nicholas FitzGerald and Keren Gu-Lemberg and Mina Khan and Lisa Anne Hendricks and Marie Pellat and Vladimir Feinberg and James Cobon-Kerr and Tara Sainath and Maribeth Rauh and Sayed Hadi Hashemi and Richard Ives and Yana Hasson and Eric Noland and Yuan Cao and Nathan Byrd and Le Hou and Qingze Wang and Thibault Sottiaux and Michela Paganini and Jean-Baptiste Lespiau and Alexandre Moufarek and Samer Hassan and Kaushik Shivakumar and Joost van Amersfoort and Amol Mandhane and Pratik Joshi and Anirudh Goyal and Matthew Tung and Andrew Brock and Hannah Sheahan and Vedant Misra and Cheng Li and Nemanja Rakićević and Mostafa Dehghani and Fangyu Liu and Sid Mittal and Junhyuk Oh and Seb Noury and Eren Sezener and Fantine Huot and Matthew Lamm and Nicola De Cao and Charlie Chen and Sidharth Mudgal and Romina Stella and Kevin Brooks and Gautam Vasudevan and Chenxi Liu and Mainak Chain and Nivedita Melinkeri and Aaron Cohen and Venus Wang and Kristie Seymore and Sergey Zubkov and Rahul Goel and Summer Yue and Sai Krishnakumaran and Brian Albert and Nate Hurley and Motoki Sano and Anhad Mohananey and Jonah Joughin and Egor Filonov and Tomasz Kępa and Yomna Eldawy and Jiawern Lim and Rahul Rishi and Shirin Badiezadegan and Taylor Bos and Jerry Chang and Sanil Jain and Sri Gayatri Sundara Padmanabhan and Subha Puttagunta and Kalpesh Krishna and Leslie Baker and Norbert Kalb and Vamsi Bedapudi and Adam Kurzrok and Shuntong Lei and Anthony Yu and Oren Litvin and Xiang Zhou and Zhichun Wu and Sam Sobell and Andrea Siciliano and Alan Papir and Robby Neale and Jonas Bragagnolo and Tej Toor and Tina Chen and Valentin Anklin and Feiran Wang and Richie Feng and Milad Gholami and Kevin Ling and Lijuan Liu and Jules Walter and Hamid Moghaddam and Arun Kishore and Jakub Adamek and Tyler Mercado and Jonathan Mallinson and Siddhinita Wandekar and Stephen Cagle and Eran Ofek and Guillermo Garrido and Clemens Lombriser and Maksim Mukha and Botu Sun and Hafeezul Rahman Mohammad and Josip Matak and Yadi Qian and Vikas Peswani and Pawel Janus and Quan Yuan and Leif Schelin and Oana David and Ankur Garg and Yifan He and Oleksii Duzhyi and Anton Älgmyr and Timothée Lottaz and Qi Li and Vikas Yadav and Luyao Xu and Alex Chinien and Rakesh Shivanna and Aleksandr Chuklin and Josie Li and Carrie Spadine and Travis Wolfe and Kareem Mohamed and Subhabrata Das and Zihang Dai and Kyle He and Daniel von Dincklage and Shyam Upadhyay and Akanksha Maurya and Luyan Chi and Sebastian Krause and Khalid Salama and Pam G Rabinovitch and Pavan Kumar Reddy M and Aarush Selvan and Mikhail Dektiarev and Golnaz Ghiasi and Erdem Guven and Himanshu Gupta and Boyi Liu and Deepak Sharma and Idan Heimlich Shtacher and Shachi Paul and Oscar Akerlund and François-Xavier Aubet and Terry Huang and Chen Zhu and Eric Zhu and Elico Teixeira and Matthew Fritze and Francesco Bertolini and Liana-Eleonora Marinescu and Martin Bölle and Dominik Paulus and Khyatti Gupta and Tejasi Latkar and Max Chang and Jason Sanders and Roopa Wilson and Xuewei Wu and Yi-Xuan Tan and Lam Nguyen Thiet and Tulsee Doshi and Sid Lall and Swaroop Mishra and Wanming Chen and Thang Luong and Seth Benjamin and Jasmine Lee and Ewa Andrejczuk and Dominik Rabiej and Vipul Ranjan and Krzysztof Styrc and Pengcheng Yin and Jon Simon and Malcolm Rose Harriott and Mudit Bansal and Alexei Robsky and Geoff Bacon and David Greene and Daniil Mirylenka and Chen Zhou and Obaid Sarvana and Abhimanyu Goyal and Samuel Andermatt and Patrick Siegler and Ben Horn and Assaf Israel and Francesco Pongetti and Chih-Wei "Louis" Chen and Marco Selvatici and Pedro Silva and Kathie Wang and Jackson Tolins and Kelvin Guu and Roey Yogev and Xiaochen Cai and Alessandro Agostini and Maulik Shah and Hung Nguyen and Noah Ó Donnaile and Sébastien Pereira and Linda Friso and Adam Stambler and Adam Kurzrok and Chenkai Kuang and Yan Romanikhin and Mark Geller and ZJ Yan and Kane Jang and Cheng-Chun Lee and Wojciech Fica and Eric Malmi and Qijun Tan and Dan Banica and Daniel Balle and Ryan Pham and Yanping Huang and Diana Avram and Hongzhi Shi and Jasjot Singh and Chris Hidey and Niharika Ahuja and Pranab Saxena and Dan Dooley and Srividya Pranavi Potharaju and Eileen O'Neill and Anand Gokulchandran and Ryan Foley and Kai Zhao and Mike Dusenberry and Yuan Liu and Pulkit Mehta and Ragha Kotikalapudi and Chalence Safranek-Shrader and Andrew Goodman and Joshua Kessinger and Eran Globen and Prateek Kolhar and Chris Gorgolewski and Ali Ibrahim and Yang Song and Ali Eichenbaum and Thomas Brovelli and Sahitya Potluri and Preethi Lahoti and Cip Baetu and Ali Ghorbani and Charles Chen and Andy Crawford and Shalini Pal and Mukund Sridhar and Petru Gurita and Asier Mujika and Igor Petrovski and Pierre-Louis Cedoz and Chenmei Li and Shiyuan Chen and Niccolò Dal Santo and Siddharth Goyal and Jitesh Punjabi and Karthik Kappaganthu and Chester Kwak and Pallavi LV and Sarmishta Velury and Himadri Choudhury and Jamie Hall and Premal Shah and Ricardo Figueira and Matt Thomas and Minjie Lu and Ting Zhou and Chintu Kumar and Thomas Jurdi and Sharat Chikkerur and Yenai Ma and Adams Yu and Soo Kwak and Victor Ähdel and Sujeevan Rajayogam and Travis Choma and Fei Liu and Aditya Barua and Colin Ji and Ji Ho Park and Vincent Hellendoorn and Alex Bailey and Taylan Bilal and Huanjie Zhou and Mehrdad Khatir and Charles Sutton and Wojciech Rzadkowski and Fiona Macintosh and Konstantin Shagin and Paul Medina and Chen Liang and Jinjing Zhou and Pararth Shah and Yingying Bi and Attila Dankovics and Shipra Banga and Sabine Lehmann and Marissa Bredesen and Zifan Lin and John Eric Hoffmann and Jonathan Lai and Raynald Chung and Kai Yang and Nihal Balani and Arthur Bražinskas and Andrei Sozanschi and Matthew Hayes and Héctor Fernández Alcalde and Peter Makarov and Will Chen and Antonio Stella and Liselotte Snijders and Michael Mandl and Ante Kärrman and Paweł Nowak and Xinyi Wu and Alex Dyck and Krishnan Vaidyanathan and Raghavender R and Jessica Mallet and Mitch Rudominer and Eric Johnston and Sushil Mittal and Akhil Udathu and Janara Christensen and Vishal Verma and Zach Irving and Andreas Santucci and Gamaleldin Elsayed and Elnaz Davoodi and Marin Georgiev and Ian Tenney and Nan Hua and Geoffrey Cideron and Edouard Leurent and Mahmoud Alnahlawi and Ionut Georgescu and Nan Wei and Ivy Zheng and Dylan Scandinaro and Heinrich Jiang and Jasper Snoek and Mukund Sundararajan and Xuezhi Wang and Zack Ontiveros and Itay Karo and Jeremy Cole and Vinu Rajashekhar and Lara Tumeh and Eyal Ben-David and Rishub Jain and Jonathan Uesato and Romina Datta and Oskar Bunyan and Shimu Wu and John Zhang and Piotr Stanczyk and Ye Zhang and David Steiner and Subhajit Naskar and Michael Azzam and Matthew Johnson and Adam Paszke and Chung-Cheng Chiu and Jaume Sanchez Elias and Afroz Mohiuddin and Faizan Muhammad and Jin Miao and Andrew Lee and Nino Vieillard and Jane Park and Jiageng Zhang and Jeff Stanway and Drew Garmon and Abhijit Karmarkar and Zhe Dong and Jong Lee and Aviral Kumar and Luowei Zhou and Jonathan Evens and William Isaac and Geoffrey Irving and Edward Loper and Michael Fink and Isha Arkatkar and Nanxin Chen and Izhak Shafran and Ivan Petrychenko and Zhe Chen and Johnson Jia and Anselm Levskaya and Zhenkai Zhu and Peter Grabowski and Yu Mao and Alberto Magni and Kaisheng Yao and Javier Snaider and Norman Casagrande and Evan Palmer and Paul Suganthan and Alfonso Castaño and Irene Giannoumis and Wooyeol Kim and Mikołaj Rybiński and Ashwin Sreevatsa and Jennifer Prendki and David Soergel and Adrian Goedeckemeyer and Willi Gierke and Mohsen Jafari and Meenu Gaba and Jeremy Wiesner and Diana Gage Wright and Yawen Wei and Harsha Vashisht and Yana Kulizhskaya and Jay Hoover and Maigo Le and Lu Li and Chimezie Iwuanyanwu and Lu Liu and Kevin Ramirez and Andrey Khorlin and Albert Cui and Tian LIN and Marcus Wu and Ricardo Aguilar and Keith Pallo and Abhishek Chakladar and Ginger Perng and Elena Allica Abellan and Mingyang Zhang and Ishita Dasgupta and Nate Kushman and Ivo Penchev and Alena Repina and Xihui Wu and Tom van der Weide and Priya Ponnapalli and Caroline Kaplan and Jiri Simsa and Shuangfeng Li and Olivier Dousse and Fan Yang and Jeff Piper and Nathan Ie and Rama Pasumarthi and Nathan Lintz and Anitha Vijayakumar and Daniel Andor and Pedro Valenzuela and Minnie Lui and Cosmin Paduraru and Daiyi Peng and Katherine Lee and Shuyuan Zhang and Somer Greene and Duc Dung Nguyen and Paula Kurylowicz and Cassidy Hardin and Lucas Dixon and Lili Janzer and Kiam Choo and Ziqiang Feng and Biao Zhang and Achintya Singhal and Dayou Du and Dan McKinnon and Natasha Antropova and Tolga Bolukbasi and Orgad Keller and David Reid and Daniel Finchelstein and Maria Abi Raad and Remi Crocker and Peter Hawkins and Robert Dadashi and Colin Gaffney and Ken Franko and Anna Bulanova and Rémi Leblond and Shirley Chung and Harry Askham and Luis C. Cobo and Kelvin Xu and Felix Fischer and Jun Xu and Christina Sorokin and Chris Alberti and Chu-Cheng Lin and Colin Evans and Alek Dimitriev and Hannah Forbes and Dylan Banarse and Zora Tung and Mark Omernick and Colton Bishop and Rachel Sterneck and Rohan Jain and Jiawei Xia and Ehsan Amid and Francesco Piccinno and Xingyu Wang and Praseem Banzal and Daniel J. Mankowitz and Alex Polozov and Victoria Krakovna and Sasha Brown and MohammadHossein Bateni and Dennis Duan and Vlad Firoiu and Meghana Thotakuri and Tom Natan and Matthieu Geist and Ser tan Girgin and Hui Li and Jiayu Ye and Ofir Roval and Reiko Tojo and Michael Kwong and James Lee-Thorp and Christopher Yew and Danila Sinopalnikov and Sabela Ramos and John Mellor and Abhishek Sharma and Kathy Wu and David Miller and Nicolas Sonnerat and Denis Vnukov and Rory Greig and Jennifer Beattie and Emily Caveness and Libin Bai and Julian Eisenschlos and Alex Korchemniy and Tomy Tsai and Mimi Jasarevic and Weize Kong and Phuong Dao and Zeyu Zheng and Frederick Liu and Fan Yang and Rui Zhu and Tian Huey Teh and Jason Sanmiya and Evgeny Gladchenko and Nejc Trdin and Daniel Toyama and Evan Rosen and Sasan Tavakkol and Linting Xue and Chen Elkind and Oliver Woodman and John Carpenter and George Papamakarios and Rupert Kemp and Sushant Kafle and Tanya Grunina and Rishika Sinha and Alice Talbert and Diane Wu and Denese Owusu-Afriyie and Cosmo Du and Chloe Thornton and Jordi Pont-Tuset and Pradyumna Narayana and Jing Li and Saaber Fatehi and John Wieting and Omar Ajmeri and Benigno Uria and Yeongil Ko and Laura Knight and Amélie Héliou and Ning Niu and Shane Gu and Chenxi Pang and Yeqing Li and Nir Levine and Ariel Stolovich and Rebeca Santamaria-Fernandez and Sonam Goenka and Wenny Yustalim and Robin Strudel and Ali Elqursh and Charlie Deck and Hyo Lee and Zonglin Li and Kyle Levin and Raphael Hoffmann and Dan Holtmann-Rice and Olivier Bachem and Sho Arora and Christy Koh and Soheil Hassas Yeganeh and Siim Põder and Mukarram Tariq and Yanhua Sun and Lucian Ionita and Mojtaba Seyedhosseini and Pouya Tafti and Zhiyu Liu and Anmol Gulati and Jasmine Liu and Xinyu Ye and Bart Chrzaszcz and Lily Wang and Nikhil Sethi and Tianrun Li and Ben Brown and Shreya Singh and Wei Fan and Aaron Parisi and Joe Stanton and Vinod Koverkathu and Christopher A. Choquette-Choo and Yunjie Li and TJ Lu and Abe Ittycheriah and Prakash Shroff and Mani Varadarajan and Sanaz Bahargam and Rob Willoughby and David Gaddy and Guillaume Desjardins and Marco Cornero and Brona Robenek and Bhavishya Mittal and Ben Albrecht and Ashish Shenoy and Fedor Moiseev and Henrik Jacobsson and Alireza Ghaffarkhah and Morgane Rivière and Alanna Walton and Clément Crepy and Alicia Parrish and Zongwei Zhou and Clement Farabet and Carey Radebaugh and Praveen Srinivasan and Claudia van der Salm and Andreas Fidjeland and Salvatore Scellato and Eri Latorre-Chimoto and Hanna Klimczak-Plucińska and David Bridson and Dario de Cesare and Tom Hudson and Piermaria Mendolicchio and Lexi Walker and Alex Morris and Matthew Mauger and Alexey Guseynov and Alison Reid and Seth Odoom and Lucia Loher and Victor Cotruta and Madhavi Yenugula and Dominik Grewe and Anastasia Petrushkina and Tom Duerig and Antonio Sanchez and Steve Yadlowsky and Amy Shen and Amir Globerson and Lynette Webb and Sahil Dua and Dong Li and Surya Bhupatiraju and Dan Hurt and Haroon Qureshi and Ananth Agarwal and Tomer Shani and Matan Eyal and Anuj Khare and Shreyas Rammohan Belle and Lei Wang and Chetan Tekur and Mihir Sanjay Kale and Jinliang Wei and Ruoxin Sang and Brennan Saeta and Tyler Liechty and Yi Sun and Yao Zhao and Stephan Lee and Pandu Nayak and Doug Fritz and Manish Reddy Vuyyuru and John Aslanides and Nidhi Vyas and Martin Wicke and Xiao Ma and Evgenii Eltyshev and Nina Martin and Hardie Cate and James Manyika and Keyvan Amiri and Yelin Kim and Xi Xiong and Kai Kang and Florian Luisier and Nilesh Tripuraneni and David Madras and Mandy Guo and Austin Waters and Oliver Wang and Joshua Ainslie and Jason Baldridge and Han Zhang and Garima Pruthi and Jakob Bauer and Feng Yang and Riham Mansour and Jason Gelman and Yang Xu and George Polovets and Ji Liu and Honglong Cai and Warren Chen and XiangHai Sheng and Emily Xue and Sherjil Ozair and Christof Angermueller and Xiaowei Li and Anoop Sinha and Weiren Wang and Julia Wiesinger and Emmanouil Koukoumidis and Yuan Tian and Anand Iyer and Madhu Gurumurthy and Mark Goldenson and Parashar Shah and MK Blake and Hongkun Yu and Anthony Urbanowicz and Jennimaria Palomaki and Chrisantha Fernando and Ken Durden and Harsh Mehta and Nikola Momchev and Elahe Rahimtoroghi and Maria Georgaki and Amit Raul and Sebastian Ruder and Morgan Redshaw and Jinhyuk Lee and Denny Zhou and Komal Jalan and Dinghua Li and Blake Hechtman and Parker Schuh and Milad Nasr and Kieran Milan and Vladimir Mikulik and Juliana Franco and Tim Green and Nam Nguyen and Joe Kelley and Aroma Mahendru and Andrea Hu and Joshua Howland and Ben Vargas and Jeffrey Hui and Kshitij Bansal and Vikram Rao and Rakesh Ghiya and Emma Wang and Ke Ye and Jean Michel Sarr and Melanie Moranski Preston and Madeleine Elish and Steve Li and Aakash Kaku and Jigar Gupta and Ice Pasupat and Da-Cheng Juan and Milan Someswar and Tejvi M. and Xinyun Chen and Aida Amini and Alex Fabrikant and Eric Chu and Xuanyi Dong and Amruta Muthal and Senaka Buthpitiya and Sarthak Jauhari and Nan Hua and Urvashi Khandelwal and Ayal Hitron and Jie Ren and Larissa Rinaldi and Shahar Drath and Avigail Dabush and Nan-Jiang Jiang and Harshal Godhia and Uli Sachs and Anthony Chen and Yicheng Fan and Hagai Taitelbaum and Hila Noga and Zhuyun Dai and James Wang and Chen Liang and Jenny Hamer and Chun-Sung Ferng and Chenel Elkind and Aviel Atias and Paulina Lee and Vít Listík and Mathias Carlen and Jan van de Kerkhof and Marcin Pikus and Krunoslav Zaher and Paul Müller and Sasha Zykova and Richard Stefanec and Vitaly Gatsko and Christoph Hirnschall and Ashwin Sethi and Xingyu Federico Xu and Chetan Ahuja and Beth Tsai and Anca Stefanoiu and Bo Feng and Keshav Dhandhania and Manish Katyal and Akshay Gupta and Atharva Parulekar and Divya Pitta and Jing Zhao and Vivaan Bhatia and Yashodha Bhavnani and Omar Alhadlaq and Xiaolin Li and Peter Danenberg and Dennis Tu and Alex Pine and Vera Filippova and Abhipso Ghosh and Ben Limonchik and Bhargava Urala and Chaitanya Krishna Lanka and Derik Clive and Yi Sun and Edward Li and Hao Wu and Kevin Hongtongsak and Ianna Li and Kalind Thakkar and Kuanysh Omarov and Kushal Majmundar and Michael Alverson and Michael Kucharski and Mohak Patel and Mudit Jain and Maksim Zabelin and Paolo Pelagatti and Rohan Kohli and Saurabh Kumar and Joseph Kim and Swetha Sankar and Vineet Shah and Lakshmi Ramachandruni and Xiangkai Zeng and Ben Bariach and Laura Weidinger and Tu Vu and Alek Andreev and Antoine He and Kevin Hui and Sheleem Kashem and Amar Subramanya and Sissie Hsiao and Demis Hassabis and Koray Kavukcuoglu and Adam Sadovsky and Quoc Le and Trevor Strohman and Yonghui Wu and Slav Petrov and Jeffrey Dean and Oriol Vinyals},
      year={2024},
      eprint={2312.11805},
      archivePrefix={arXiv},
      primaryClass={cs.CL},
      url={https://arxiv.org/abs/2312.11805}, 
}

@misc{zhang2022optopenpretrainedtransformer,
      title={OPT: Open Pre-trained Transformer Language Models}, 
      author={Susan Zhang and Stephen Roller and Naman Goyal and Mikel Artetxe and Moya Chen and Shuohui Chen and Christopher Dewan and Mona Diab and Xian Li and Xi Victoria Lin and Todor Mihaylov and Myle Ott and Sam Shleifer and Kurt Shuster and Daniel Simig and Punit Singh Koura and Anjali Sridhar and Tianlu Wang and Luke Zettlemoyer},
      year={2022},
      eprint={2205.01068},
      archivePrefix={arXiv},
      primaryClass={cs.CL},
      url={https://arxiv.org/abs/2205.01068}, 
}

@misc{bellagente2024stablelm216b,
      title={Stable LM 2 1.6B Technical Report}, 
      author={Marco Bellagente and Jonathan Tow and Dakota Mahan and Duy Phung and Maksym Zhuravinskyi and Reshinth Adithyan and James Baicoianu and Ben Brooks and Nathan Cooper and Ashish Datta and Meng Lee and Emad Mostaque and Michael Pieler and Nikhil Pinnaparju and Paulo Rocha and Harry Saini and Hannah Teufel and Niccolo Zanichelli and Carlos Riquelme},
      year={2024},
      eprint={2402.17834},
      archivePrefix={arXiv},
      primaryClass={cs.CL},
      url={https://arxiv.org/abs/2402.17834}, 
}

\end{document}